\documentclass[twocolumn]{aastex63}

\usepackage{graphicx}
\usepackage{dcolumn}
\usepackage{bm}

\usepackage{makecell}
\usepackage{verbatim} 
\usepackage{amsmath}

\begin{document}


\title{A Significant Increase in Detection of High-Resolution Emission Spectra Using a Three-Dimensional Atmospheric Model of a Hot Jupiter}


\correspondingauthor{Hayley Beltz}
 \email{hbeltz@umich.edu}

\author[0000-0002-6980-052X]{Hayley Beltz}

 \affiliation{Department of Astronomy, University of Michigan, Ann Arbor, MI 48109, USA}
\author[0000-0003-3963-9672]{Emily Rauscher}
\affiliation{Department of Astronomy, University of Michigan, Ann Arbor, MI 48109, USA}

\author[0000-0002-7704-0153]{Matteo Brogi}
    \affiliation{Department of Physics, University of Warwick, Coventry CV4 7AL, UK}
    \affiliation{INAF - Osservatorio Astrofisico di Torino, Via Osservatorio 20, I-10025 Pino Torinese, Italy}
    \affiliation{Centre for Exoplanets and Habitability, University of Warwick, Gibbet Hill Road, Coventry CV4 7AL, UK}
    
\author[0000-0002-1337-9051]{Eliza M.-R.\ Kempton}
\affil{Department of Astronomy, University of Maryland, College Park, MD 20742, USA}

\begin{abstract}

High resolution spectroscopy has opened the way for new, detailed study of exoplanet atmospheres. There is evidence that this technique can be sensitive to the complex, three-dimensional (3D) atmospheric structure of these planets. In this work, we perform cross correlation analysis on high resolution ($R\sim100,000$) CRIRES/VLT emission spectra of the Hot Jupiter HD~209458b. We generate template emission spectra from a 3D atmospheric circulation model of the planet, accounting for temperature structure and atmospheric motions \added{---winds and  planetary rotation---} missed by spectra calculated from one-dimensional models. In this first-of-its-kind analysis, we find that using \added{template spectra generated from}a 3D model produces a more significant detection (\replaced{6.8}{6.9} $\sigma$) of the planet's signal than any of the hundreds of one-dimensional models we tested (maximum of $5.1 \sigma$). We recover the planet's thermal emission, its orbital motion, and the presence of CO in its atmosphere at high significance.  Additionally, we analyzed the relative influences of 3D temperature and chemical structures in this improved detection, including the contributions from CO and H$_{2}$O, as well as the role of atmospheric Doppler signatures from winds and rotation. This work shows that the Hot Jupiter's 3D atmospheric structure has a first-order influence on its emission spectra at high resolution and motivates the use of \replaced{complex}{multi-dimensional} atmospheric models in high-resolution spectral analysis.

\vspace{0.4in}
\end{abstract}

\section{Introduction}

High Resolution Spectroscopy (HRS) is a relatively recent, powerful method for exoplanet atmospheric characterization. It uses a spectral resolution high enough ($R \gtrsim 30,000$) to unambiguously detect the unique sets of spectral lines from atoms or molecules in an exoplanet's spectrum. While the planet's spectrum is often  orders of magnitude weaker than the stellar noise, its signal can be detected via cross-correlation with a template spectrum \added{due to the increased number of lines present at high resolution.}. This is accomplished by exploiting the planet's changing orbital radial velocity along the observers line of sight  which helps to remove the stellar and telluric signals, whose spectral features remain effectively at fixed wavelengths over the duration of a typical observation. \added{By removing the components of spectrum that are constant with time, one is left with noise and the planet spectrum, which can then be detected via cross-correlation.} \citet{birkbyhires} presents a review of the HRS method and recent results from its use. 

 HRS was first applied to the well-known hot Jupiter HD~209458b using the CRIRES instrument on the VLT \citep{2010Natur.465.1049S}, definitively detecting CO in transmission spectra from the planet. Further analysis of the transmission spectra of this planet at high resolution have resulted in detections of water vapor \citep{sanchezlopez2019} and helium \citep{alonso2019}. Emission spectra of this planet have also been measured with HRS, providing evidence against an atmospheric temperature inversion \citep{schwarz}, as well as  determining both carbon monoxide and water abundances when combined with lower resolution data \citep{combininglowandhigh,combinedhudra}. In this paper we present a re-analysis of the previously published CRIRES/VLT data for this planet \citep{schwarz}, but with template spectra generated from a three-dimensional atmospheric model.

One of the unique strengths of HRS is that at the highest resolutions ($R \sim 100,000$) the observed spectra can contain information about the \textit{atmospheric} motion of the planet. The original HRS result by \citet{2010Natur.465.1049S} found hints of day-to-night winds on the planet in a net blue-shift of the planet's spectrum by $2 \pm 1$ km s$^{-1}$ (during transit, day-to-night winds blow toward the observer). Transmission spectra of the hot Jupiter HD~189733b also show evidence for atmospheric motion, including both net Doppler shifts from winds and Doppler broadening from a combination of rotation and eastward equatorial winds \citep{louden2015,Brogianalysis,Flowers}. Measured Doppler broadening in high-resolution emission spectra of directly imaged planets/companions have also been used to constrain the rotation rates of these objects \citep{rotationbpic,Schwarz2016,Bryan2020}.

The two sources of atmospheric motion---winds and rotation---are not physically independent.\added{For a recent review of hot Jupiter dynamics, see \citet{showman2020review}.} One of the governing forces in determining atmospheric circulation is the Coriolis force,  meaning that the rotation rate of a planet strongly influences the wind structure and speeds. Hot Jupiters are commonly assumed to be tidally locked into rotation rates synchronous with their orbits \citep[e.g.,][]{Rasio1996}, but deviations from this expected rotation state would have consequences for the speed and structure of atmospheric winds \citep{Showman2009}, which then influences the expected Doppler shifts and broadening in HRS data \citep{Rauscher_rotationrates}. It is an ongoing debate within the community as to how tidal forces interact with the complex structure of hot Jupiters and whether we should assume them to be synchronized or not \citep{Gu2009,Arras2010,Auclair2018,Lee2020,Yu2020}.

Given the exquisite spectral detail measurable by HRS, including constraints on atmospheric motions, we may wonder how sensitive it is to the full three-dimensional nature of the planet; and what degree of bias will a one dimensional model introduce.  Another way to state this is whether or not one-dimensional atmospheric models are sufficient to accurately interpret HRS data. Especially for hot Jupiters, where we expect hundreds of Kelvin temperature contrasts across the globe \added{\citep{GCM, DobbsDixon2013,kataria2016,Parmentier2018,Deitrick2020,Drummond2020} } , differences in the local atmospheric structure can result in limb- or disk-integrated transmission or emission spectra (respectively) that are significantly different from a spectrum calculated using a 1-D model. Several studies have considered how the 3-D nature of a planet can influence lower resolution spectra \citep[e.g.,][]{Fortney2006,Fortney2010,Burrows2010} and, in particular, ways that the use of 1-D models could bias our interpretation of spectral data \citep[e.g.,][]{Feng2016,Blecic2017,Caldas2019,Pluriel2020}. For HRS data, several studies have simulated high-resolution spectra from different 3-D models, both in transmission \citep{Kempton2012,showman2013,Kempton2014,Rauscher_rotationrates} and emission \citep{jisheng,Harada2019}, demonstrating that the complex atmospheric structures of hot Jupiters can influence HRS data. 

\citet{Flowers} presented a first-of-its-kind analysis of HRS data, using simulated transmission spectra from 3-D models as template spectra in the cross-correlation analysis of observations of the hot Jupiter HD~189733b. Not only was the planet's signal detected at high significance (supporting the validity of the 3-D models), but this work also consistently detected the Doppler signature of day-to-night winds on this planet. \replaced{This was accomplished by showing that the best signal detection required an anomalous Doppler shift of the planet if the effect of the winds was artificially turned off in the simulated spectra, but had correct planetary motion with the Doppler effects of the winds included.}{When the Doppler effects from the winds were artificially excluded from the calculation of the template spectra, the planet's signal was detected with an anomalous blue-shift; when the effects of the winds were included, the detection was at the expected planet velocity.} \added{That is, ignoring the Doppler effects on simulated transmission spectra resulted in incorrect inferred planetary motion, confirming their measurable influence in the observed spectra.  }

Here we present an analogous study to \citet{Flowers},  but for emission spectra (as opposed to transmission), in which we use simulated spectra from 3-D models in the HRS cross-correlation analysis. In addition to studying a complimentary observational technique---emission instead of transmission---we also target a different bright hot Jupiter than that analysis, namely HD~209458b. HRS transmission spectra can be directly influenced by atmospheric motion, but are only secondarily affected by the three-dimensional temperature structure \added{ \citet{Flowers}}. We expect that HRS emission spectra may be much more sensitive to differences in atmospheric \added{thermal} structure around the planet, given that any Doppler effects from atmospheric motion will be most sensitive to the brightest regions of the planet \citep{jisheng}. In this paper we empirically determine how sensitive HRS emission spectra are to the 3-D nature of a particular planet, as well as to what degree various aspects of the atmospheric structure contribute to the observed data. Specifically, we study the sensitivity of the data to the planet's rotation period by running a suite of 3-D models for a range of rotation rates, producing a set of consistent temperature and wind structures for each case. We also test the sensitivity of the data to atmospheric chemistry by comparing an assumption of well-mixed abundances or local chemical equilibrium values in the radiative transfer routine we use to post-process the 3-D models and create simulated spectra. We also analyze the relative contributions of the two main opacity sources over the wavelengths of observation (2.285 to 2.348 $\mu$m):
carbon monoxide and water. Finally, we test the sensitivity of the data to Doppler effects from atmospheric motions by cross-correlating with simulated spectra calculated with and without those effects. 

In Section \ref{sec:numerical}, we explain the various numerical methods used in this work: the three-dimensional atmospheric model and the radiative transfer routine used to post-process the 3-D models and calculate simulated emission spectra. Additionally, we briefly describe the results of these standard hot Jupiter models. In Section \ref{sec:data} we describe the observational data, along with details of our reduction and analysis methods. In Section \ref{sec:CC} we present the results of our cross-correlation analysis, comparing the strength of planetary signal detected when using template spectra from 1-D or 3-D models, and comparing the aforementioned assumptions regarding chemistry, opacity sources, Doppler effects, and rotation rates. In Section \ref{sec:conclusion} we summarize our main results.

\section{Numerical Models: 3D GCMs and Simulated Emission Spectra} \label{sec:numerical}
In order to create simulated high-resolution emission spectra for HD~209458b, we first use a General Circulation Model to predict the three-dimensional atmospheric structure of the planet---that is, thermal and wind structure--- and then post-process the results using a detailed radiative transfer routine that accounts for the correct geometry and atmospheric Doppler shifts. These modeling methods and results are not particularly novel, having formed the basis of previous papers \citep{Kempton2012,GCM,Rauscher_rotationrates,newrad,jisheng}; however, our suite of models for this particular planet have not been published previously and so we briefly describe the results in order to set the stage for the comparison between the simulated emission spectra and observed data.

\subsection{General Circulation Model} \label{sec:gcm}
General Circulation Models (GCMs) are three-dimensional computational atmospheric models that simulate the underlying physics and circulation patterns of planetary atmospheres. For this work, we utilized the GCM  from \cite{GCM} with the radiative transfer scheme upgraded as described in \cite{newrad}. This model solves the primitive equations of meteorology: the standard set of fluid dynamics equations with simplifying assumptions appropriate for the atmospheric context, solved in the rotating frame of the planet \citep[see an early review by][]{SCM2010}.
The radiative heating and cooling of the atmospheric uses a double-gray scheme. That is, radiation is treated with two different absorption coefficients under two regimes; 
an infrared coefficient to model the thermal interaction of the gas with radiation and an optical coefficient to model the absorption of incoming starlight. For a more detailed explanation of the GCM, see \citet{GCM} and \citet{newrad}. 
We model the hot Jupiter HD~209458b using the parameters listed in Table \ref{tab:gcm_params}, with system parameters from \citet{hd209params}, a high internal heat flux appropriate for this inflated hot Jupiter \citep{Thorngren2019}, and absorption coefficients and gas properties set to match our previous models of hot Jupiter atmospheres \citep[e.g.,][]{GCM}. Typically, we assume that hot Jupiters have been tidally locked into synchronous orbits, \added{meaning that the rotation period and orbital period are equal}. In order to empirically test this, we ran the GCM for a total of 12 different rotation rates spanning values faster and slower than synchronous. \added{The slowest rotation rate was chosen to ensure that at least one of the models fell into the disrupted circulation regime for slow rotation previously found in \citet{Rauscher_rotationrates}. We then extended our rotation rate sampling (at 0.25 km/s in rotation speed) to comparably cover faster rotation rates.} We list the set of chosen rotation periods and their corresponding equatorial rotational velocities in Table \ref{tab:rot_models}, along with some representative wind speeds from each model.

We ran each model at a horizontal spectral resolution of T31, corresponding to a physical scale of $\sim$4 degrees at the equator and with 45 vertical layers evenly spaced in log pressure from 100 bar to 10 microbar. The planets were initialized with a globally averaged temperature-pressure profile and no winds.\added{See \citet{Guillot2010} for a derivation of profiles and \citet{RauscherandKempton2014} for a discussion of the global averaging parameter chosen (set to $f=0.375$ here). } Each simulation was allowed to run for 3000 orbits; by this point the upper atmosphere (including the infrared photosphere) had reached a steady state. \citet{Carone2019} recently demonstrated that the treatment of the deep atmosphere in hot Jupiter simulations---in particular the depth of the bottom boundary and the assumed strengths of convective adjustment and frictional/magnetic damping---can influence the circulation results predicted for the upper, observable atmosphere. Nevertheless, their models of HD~209458b show that this planet exhibits the standard hot Jupiter circulation pattern, in agreement with our results here.

\begin{deluxetable}{lc}
\caption{HD 209458b System Parameters} 
\label{tab:gcm_params}
\tablehead{ \colhead{Parameter} & \colhead{Value}} 
\startdata
         Planet radius, $R_{p}$ & $9.9 \times 10^{7}$ m \\
         Gravitational acceleration, $g$ & 9.434 m s$^{-2}$ \\
         \added{Orbital Period & 3.525 days} \\
         Orbital revolution rate, $\omega_{\mathrm{orb}}$ & $2.06318 \times 10^{-5}$ s$^{-1}$ \\
         \added{Synchronous rotation speed} \tablenotemark{a} & \added{2.04 km s$^{-1}$} \\
         Substellar irradiation, $F_{\mathrm{irr}}$ & $1.06 \times 10^{6}$ W m$^{-2}$\\
         Planet internal heat flux, $F_{\mathrm{int}}$ & 3500 W m$^{-2}$\\
         Optical absorption coefficient, $\kappa_{vis}$ & $4 \times 10^{-3}$ cm$^{2}$ g$^{-1}$ \\
         Infrared absorption coefficient, $\kappa_{IR}$ & $1 \times 10^{-2}$ cm $^{2}$ g$^{-1}$ \\
         Specific gas constant, $R$ & 3523 J kg$^{-1}$ K$^{-1} $\\
         Ratio of gas constant to heat capacity, $R/c_{p}$ & 0.286 \\
         Stellar radius, $R_{*}$ & $1.19 $ $ M_{\odot}$   \\
         Stellar effective temperature, $T_{*eff}$  & $6090$ K   \\
\enddata
\tablenotetext{a}{In the case of synchronous rotation, this is the corresponding \\  velocity at the equator, calculated as $2 \pi R_{p}/ \omega_{orb}$.}
\end{deluxetable}

\begin{deluxetable}{cccc} 
\caption{Suite of General Circulation Models}\label{tab:rot_models}
\tablehead{
	\colhead{Rotation} & \colhead{Rotational} & \colhead{Max.\ wind speed} &
	\colhead{Max.\ wind speed} \\
	\colhead{period} & \colhead{speed} & \colhead{at IR photosphere} & \colhead{at 0.1 mbar} \\
	\colhead{(days)} & \colhead{(km/s)} & \colhead{(km/s)} & \colhead{(km/s)}
	}
\startdata
9.08 & 0.79 & 2.50 & 6.28 \\
6.91 & 1.04 & 2.64 & 4.44 \\
5.57 & 1.29 & 5.65 & 6.87 \\
4.67 & 1.54 & 5.64 & 6.76 \\
4.02 & 1.79 & 5.61 & 6.64 \\
\textbf{3.53} & \textbf{2.04} & \textbf{5.64} & \textbf{6.32}\\
3.14 & 2.29 & 5.43 & 6.19 \\
2.83 & 2.54 & 5.47 & 6.15 \\
2.58 & 2.79 & 5.10 & 5.72 \\
2.37 & 3.04 & 4.78 & 5.56 \\
2.19 & 3.29 & 3.77 & 5.02 \\
2.03 & 3.54 & 4.62 & 5.17 \\
\enddata
\tablecomments{The bolded values are for the model in a tidally-locked, synchronous rotation state. The rotational speeds are calculated as $2\pi R_p/\omega_{\mathrm{rot}}$. Continuum emission comes from the IR photosphere (at $\sim$65 mbar), while the absorption line cores come from pressure regions nearer to 0.1 mbar. Wind speeds are measured in the rotating frame of the planet.}
\end{deluxetable}

\subsection{GCM Results} \label{sec:gcm_results}

\begin{figure}
    \centering
    \includegraphics[width=3.4in]{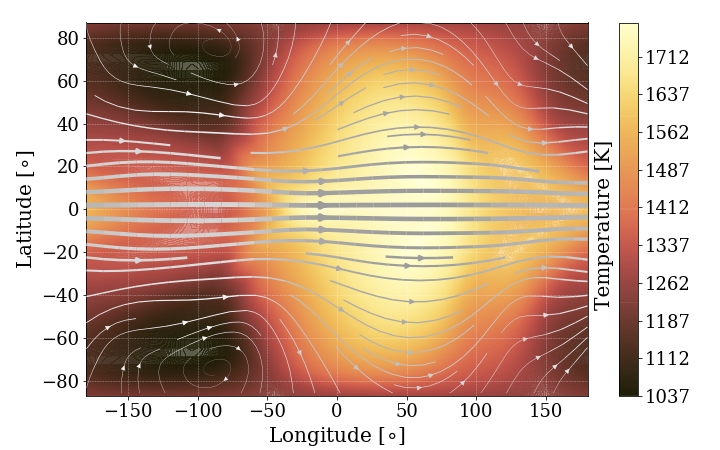}
    \caption{The temperature structure near the infrared photosphere ( $\sim$ 65 mbar), for our synchronously rotating model of HD~209458b, centered on the substellar point (at 0,0). Streamlines have been overplotted, with thicker lines showing stronger winds. In the eastward direction, the winds reach a speed of 5.6 km/s. The hottest gas has been advected to the east of the substellar point by a strong equatorial jet, in the typical hot Jupiter circulation pattern. }
    \label{fig:cylinmap}
\end{figure}

\begin{figure}
    \centering
    \includegraphics[width=3.25in]{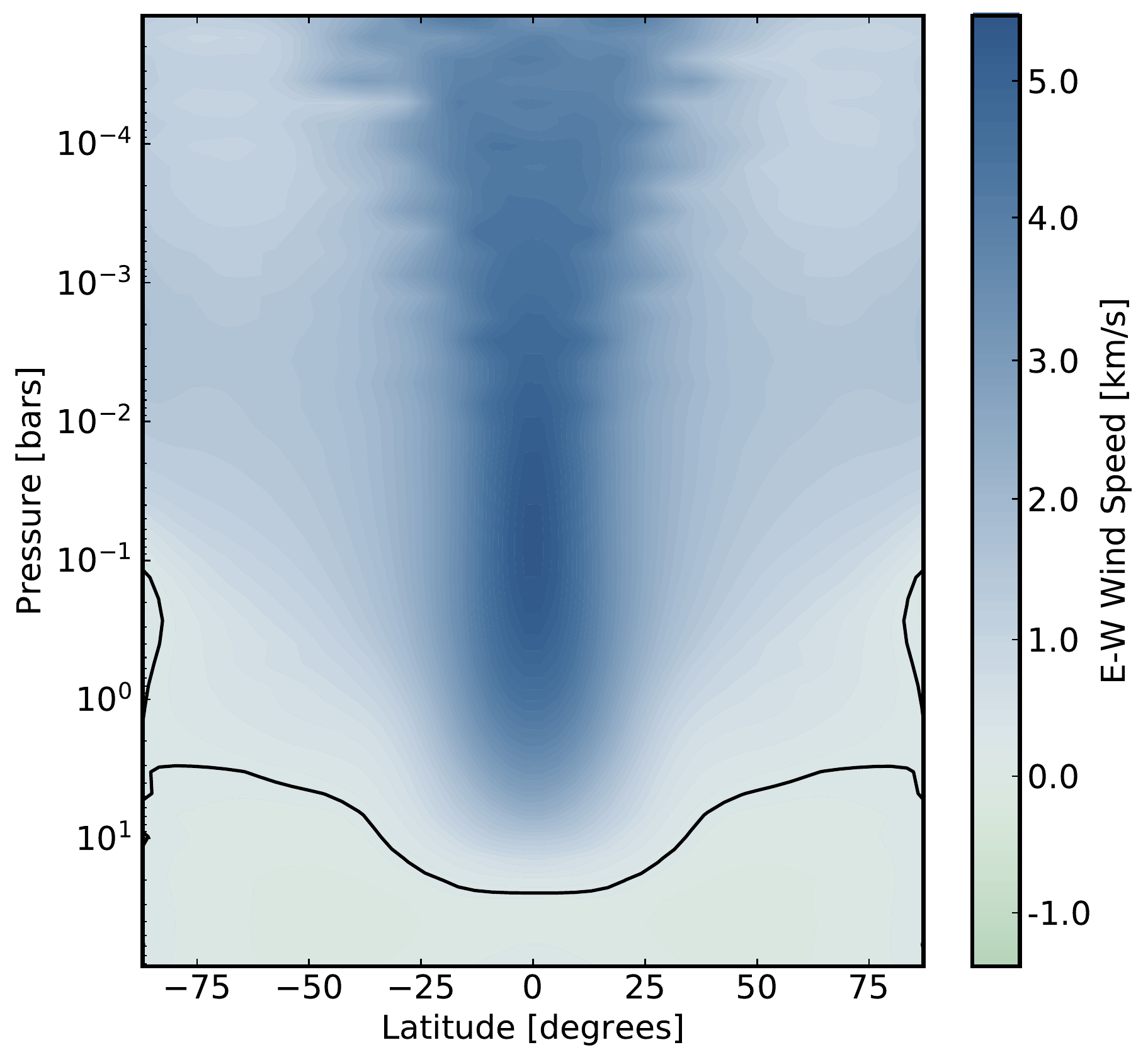}
    \caption{Longitudinally averaged east-west wind speeds throughout the atmosphere, for the synchronous rotation case. The eastward equatorial jet (dark blue) 
    extends deep into the atmosphere. The black contour shows the boundary between eastward (positive) and westward (negative) winds.} 
    \label{fig: windpressure}
\end{figure}

Most of our models display the quintessential features expected for hot Jupiters: a strong eastward equatorial jet which advects the hottest spot on the planet slightly eastward of the substellar point and reduces---but does not eliminate---a large day-to-night temperature contrast of hundreds of Kelvin. We show this temperature structure for the synchronous model in Figure \ref{fig:cylinmap}. The equatorial jet characteristically extends throughout most of the atmosphere; Figure \ref{fig: windpressure} shows the zonally averaged winds for the synchronous model. Higher in the atmosphere an additional, significant component of the winds is a substellar-to-antistellar flow pattern; in Figure \ref{fig: windpressure} this shows up as a decrease in the averaged east-west wind speed. 

Figures \ref{fig:alltemps} and \ref{fig:windgrid} in the Appendix show maps of the temperature and winds at the infrared photosphere for all of the 12 models with different rotation rates. In line with results from previous investigations of non-synchronously rotating hot Jupiters \citep{Showman2009,Rauscher_rotationrates,Flowers}, we find that as the rotation rate increases, the stronger Coriolis force causes the equatorial jet to become more narrow and eventually secondary, higher latitude jets form. The wind speeds tend to decrease with increasing rotation rate (see Table \ref{tab:rot_models}), conspiring to create generally similar temperature patterns at the infrared photospheres of each model (Figure \ref{fig:alltemps}).

\begin{figure}
    \centering
    \includegraphics[width=3.4in]{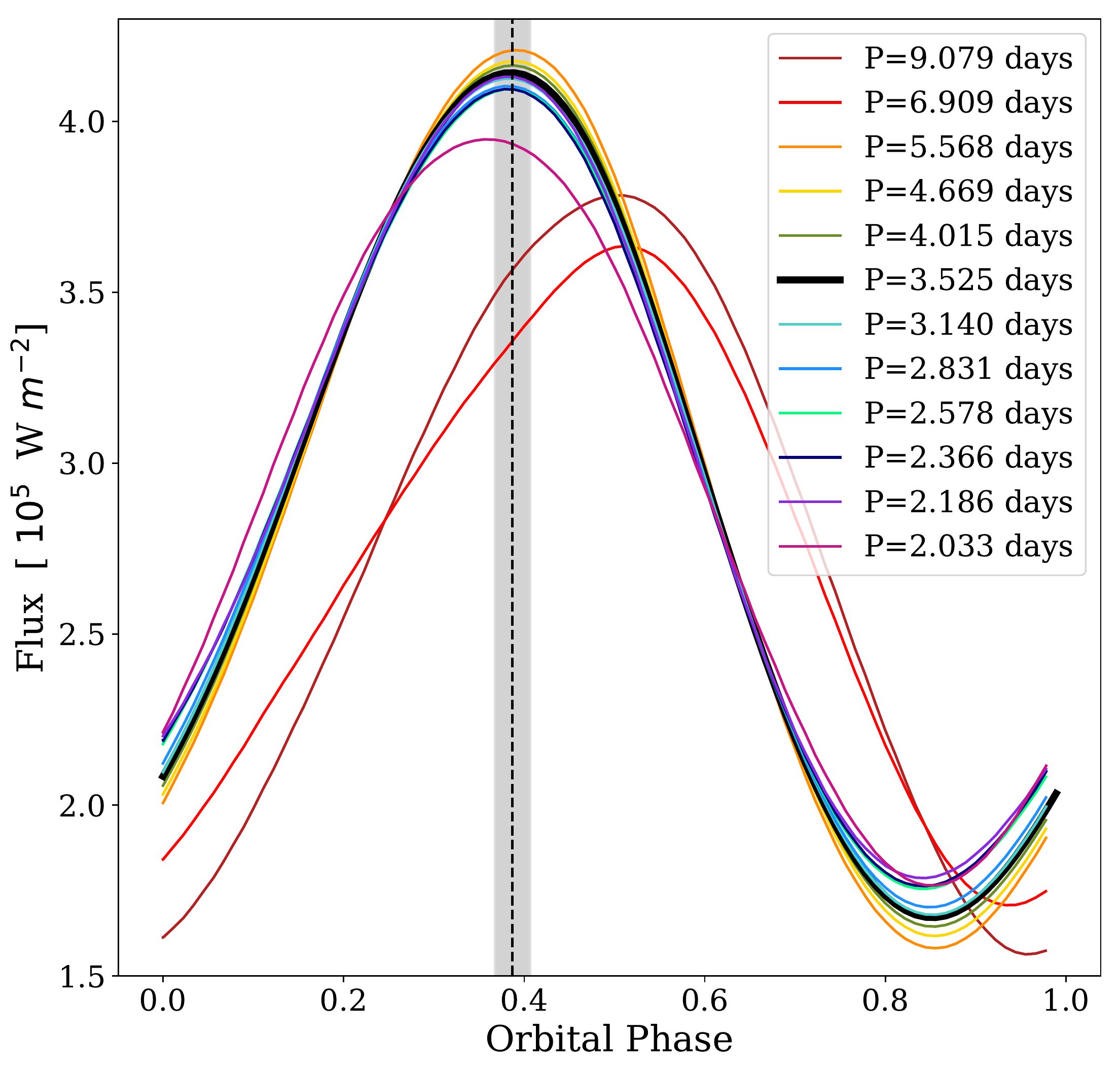}
    \caption{Calculated orbital phase curves of total thermal emission from our suite of models with different rotation rates. Only the models with the slowest two rotation rates---with circulation patterns disrupted from the standard hot Jupiter eastward flow---have phase curves that peak after secondary eclipse (which would occur at a phase of 0.5, not shown here).
    Since phase curves measured at 4.5 microns of HD~209458b show a peak before the secondary eclipse \citep[at $0.387\pm 0.017$][shown by the black dashed line and grey shaded area]{hd209phase}, we find that all models except the two slowest rotators are consistent with observations.}
    \label{fig:pcurve}
\end{figure}

The exceptions to these trends are the two most slowly rotating models, whose circulations have been disrupted from the standard hot Jupiter pattern. This disruption for very slow rotators was first identified by \citet{Rauscher_rotationrates}, and the dynamics have been studied by \citet{Penn2017}. For the purpose of this paper, these most slowly rotating models help to provide a lower limit to the possible rotation rate of HD~209458b, as the westward flow and corresponding advection of the hottest region of the atmosphere would result in an orbital phase curve of thermal emission significantly different from what has been previously observed for this planet. In Figure \ref{fig:pcurve} we show phase curves of the total thermal emission\footnote{ \added{Due to the double-gray radiative transfer in our GCM, the thermal emission is effectively bolometric, making it challenging to compare directly to the 4.5 micron flux from \citet{hd209phase}. The phase of peak flux, however, is more directly comparable as it is indicative of the photospheric temperature structure.}} from each model, calculated throughout one orbit. While most of the models do show similar curves, which peak near the measured phase of maximum flux at 4.5 micron \citep[$0.387\pm 0.017$;][]{hd209phase}, the two most slowly rotating models are ruled out by this data as they peak later in phase. Nevertheless, we include these models in the rest of our analysis in order to investigate how they are constrained by HRS data. 

\begin{figure}
    \centering
    \includegraphics[width=3.5in]{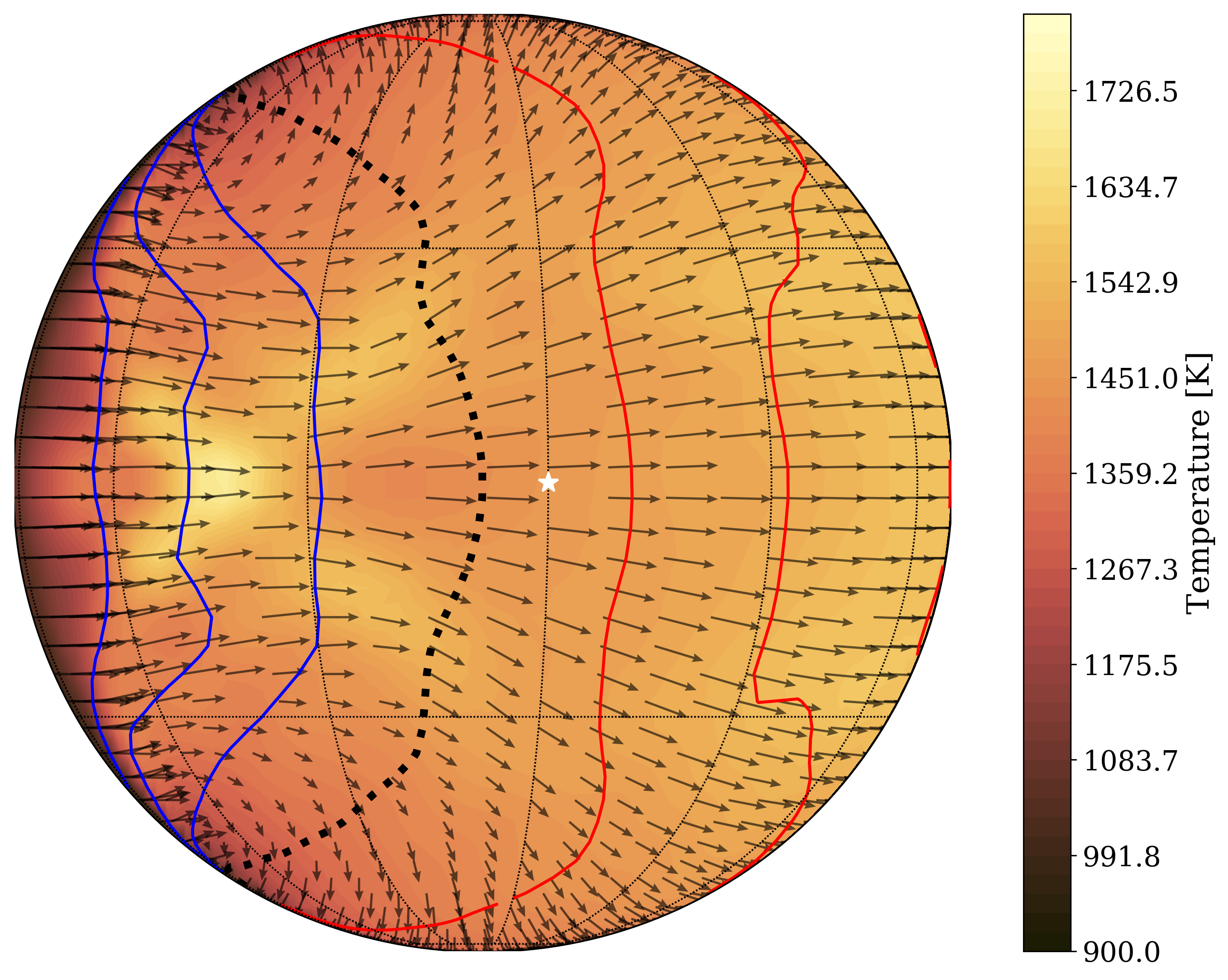}
    \caption{The temperature structure of the upper atmosphere ($\sim 0.1$ mbar), within the region from which the flux in the CO line cores emerges. The projection is centered on the subobserver point at a phase corresponding to the beginning of the observation, shortly after secondary eclipse. The substellar point is marked with a white star. Also shown are contours of the line-of-sight velocity toward (blue) or away (red) from the observer, due to contributions from both the winds and rotation of the planet (Equation \ref{vloseq}). The contour levels shown in red and blue are  $\pm$2, 4, and 6 km/s. The black dotted contour shows the boundary of 0 km/s. \added{Note that whereas at the infrared photosphere the hottest region is east of the substellar point, here it is west of the substellar point due to convergence in the atmospheric flow at that location.  } } 
    \label{fig:syncortho}
\end{figure}

Finally, since the CRIRES/VLT emission spectra of HD~209458b are the focus of our paper, we also show the temperature structure and line-of-sight velocities (from both winds and rotation) in the upper atmosphere of the synchronous model in Figure \ref{fig:syncortho}, shown in an orientation corresponding to the first night of observation.  
This is the region of the atmosphere from which the flux in the CO line cores emerges, meaning that the detailed structure of those line shapes comes from the brightness-weighted local Doppler shifts, integrated across the visible hemisphere. Since the winds are dominantly eastward, they contribute to the Doppler shifts in the same direction as the rotation field. However, the line-of-sight velocity contours are slightly bent away from being strictly aligned with the rotation axis by the specific atmospheric flow pattern.

The full set of orthographic projections for our suite of 12 models is shown in Figure \ref{fig:orthogrid} in the Appendix. Aside from the two most slowly rotating models, we see similar temperature and line of sight velocity fields across the rest of the models.  While higher blue-shifted line-of-sight velocities occur on the visible hemisphere, the red-shifted flow extends across a larger fraction of the planet disk. In contrast, the two slowest rotators have weak contributions to the velocity field from their rotation, and the winds generally work in an opposite direction to the rotation, leading to very little Doppler shifting compared to the other models. In addition, the temperature structure is fairly uniform across the visible hemisphere.

A hot feature exists on the western side of the planet (from our perspective, to the left of the subobserver point), where we also see strongly blue-shifted velocities from the combination of rotation and winds blowing around from the night side. \added{Chevron features like this, regions of flow convergence and associated heating, are commonly seen in hot Jupiter GCMs \citep[e.g.,][]{Showman2009,Rauscher2010,Komacek2019} and are related to the transport of momentum from higher latitudes to the equator \citep{ShowmanPolvani2011}. Depending on the particular model---and the pressure level within the atmosphere---chevron features may appear to the east or west of the substellar point. Here we see multiple chevron features, both near the infrared photosphere and in the upper atmosphere. New state-of-the-art GCMs in \citet{Deitrick2020} also show these features, at multiple resolutions and robust against assumptions regarding vertical hydrostatic equilibrium (see their Figures 19 and 22).}

While we have already determined that the phase curve data for HD~209458b excludes the two slowest rotation states for this planet (Figure \ref{fig:pcurve}), we are still interested to compare the simulated high-resolution emission spectra from these models to the rest of the suite. For most of the models, based on Figures \ref{fig:syncortho} and \ref{fig:orthogrid} we expect that the integrated emission spectra should show both red- and blue-shifting of the CO lines, but the detailed line shapes will be controlled by the complex three-dimensionality of the atmospheric temperature and line-of-sight velocity structures. Due to the slowing of the winds with increasing rotation rate (see Table \ref{tab:rot_models} and Figure \ref{fig:orthogrid}) we may expect similar Doppler-induced line profiles for these models. In contrast, for the two most slowly rotating models there may be very minimal Doppler effects shaping the line shapes in their simulated emission spectra.

\subsection{Radiative Transfer Post-Processing} \label{sec:rt}

\begin{figure}
    \centering
    \includegraphics[width=\columnwidth]{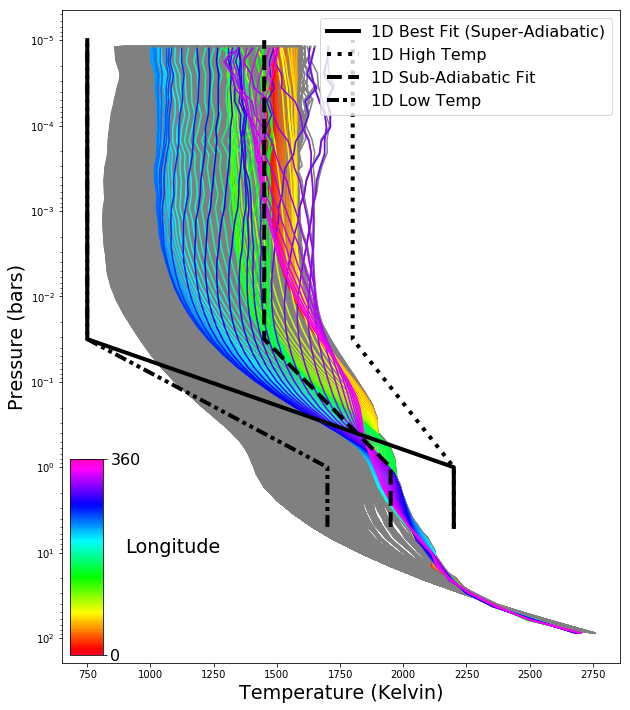}
    \caption{Temperature-pressure profiles throughout the atmosphere for our synchronously rotating model of HD~209458b. The rainbow lines show equatorial profiles, with the hue corresponding to the longitude east of the substellar point. The gray profiles are from the entire planet. We use this 3-D atmospheric structure, together with the local wind velocities, to calculate simulated high-resolution emission spectra for cross-correlation with the observed data.The black lines show examples of four temperature-pressure profiles from a suite of 1D models \added{(described in Section \ref{1Dmodels})} also  used to simulate spectra. These models cover the same temperature range realized by our 3-D models, but use only a single profile to represent the entire planet. Note that the best-fit model from this suite has an unrealistic super-adiabatic profile.}
    \label{fig: tpprof}
\end{figure}

In order to generate high-resolution emission spectra from our three-dimensional models, we apply the code and method outlined in \citet{jisheng}. Briefly, we take the output from the GCM (temperature and winds at 48 $\times$ 96 $\times$ 60 points in latitude $\times$ longitude $\times$ pressure; see Figure \ref{fig: tpprof} for the synchronous case and Figure~\ref{fig:alltp} for all of the GCM outputs) and solve the radiative transfer equation in a geometrically-consistent manner to produce the thermal emission spectrum emanating from the visible hemisphere of the planet.

The radiative transfer equation is solved in the limit of pure thermal emission: 
\begin{equation}
    I(\lambda)=B_{o}e^{-\tau_{0}}+ \int_{0}^{\tau_{o}}e^{-\tau}B\ d\tau,
\end{equation}
where $I$ is the intensity at each wavelength $\lambda$, $B$ is the Planck function (calculated from the local temperatures) and $\tau$ is the slant optical depth along the line of sight toward the observer, taking into account varying opacities throughout the path.  We strike 2,304 ($= 48 \times 96 / 2$) individual line-of-sight intensity rays through the atmosphere and then integrate with respect to the solid angle subtended by each grid cell to produce the planet's emission spectrum in flux units.  

To correctly account for the line-of-sight geometry we must first interpolate the temperature and wind output from the GCM onto a fixed-\textit{altitude} vertical grid.  This interpolation allows us to readily strike straight-through rays along the observer's sight line. This geometrically-consistent approach to the radiative transfer is somewhat unique in calculations of emission spectra from GCMs. A more common and computationally less challenging technique is to calculate the radiative transfer along radial profiles and assume isotropic emission from the top of the atmosphere. \citet{Caldas2019} have recently shown that using correct ray-tracing geometry is important in calculating transmission spectra from 3-D models; we are not aware of a similar study of geometry's importance in calculating emission spectra.\\
As a consequence of having varying temperature conditions over the visible hemisphere of the planet, we may expect that our integrated spectra are influenced by spatial variations in the chemical abundances of our main opacity sources.  Based on the temperature range spanned by the GCM outputs and the wavelength range modeled (2.28 -- 2.35 $\mu$m), we expect that H$_2$O and CO will be the dominant opacity sources.  One of the simplest assumptions we can make about the abundances of H$_2$O and CO is that they are in chemical equilibrium for the local conditions at each location in the atmosphere. However, this neglects the important influence of mixing from atmospheric dynamics, which is likely to bring these species out of chemical equilibrium. The physically and chemically sophisticated work by \citet{Drummond2020} demonstrated that 3-D mixing is expected to alter the chemical structure of hot Jupiter atmospheres, with the vertical and horizontal advection components both being significant \citep[with similar results also found by][]{Mendoca2018}. In their model of HD~209458b, however, they found minimal differences in the abundances of CO and H$_2$O between their kinetics model and the assumption of chemical equilibrium. While they predicted minimal differences between these cases in their simulated (lower resolution) emission spectra, here we further investigate the influence of chemical abundances in high resolution emission spectra.\\
The double-gray radiative transfer scheme within our GCM simplifies the multi-wavelength opacities of the atmosphere, meaning that we do not prescribe a specific chemistry, nor does the simulation predict chemical mixing. In order to investigate the impact of chemistry on the emission spectra, we consider two extreme cases within our post-processing framework: abundances determined everywhere by local chemical equilibrium, or abundances that are fully homogenized throughout the atmosphere and set to some constant volume mixing ratio (VMR). The first assumption applies to the limit where dynamics do not create any significant chemical disequilibrium, while the second may be a proxy for fully efficient mixing, with the caveat that we still need to choose a value for the homogenized abundances. We choose to fix the values for water and CO to the best-fit values from a previous retrieval analysis of these same data \citep[VMR values of $1\times 10^{-3.5}$ for CO and $1\times 10^{-5}$ for water,][]{combininglowandhigh}.  

There is significant evidence in the literature suggesting a water abundance below the solar equilibrium value \added{\citep[which would be $\sim 5 \times 10^{-4} $; ][]{h2oabundance_madhusudhan}} for HD~209458b \citep[although see][]{Line2016}. From previous analysis of these HRS data, a marginal evidence for H$_2$O was claimed by \citet{Brogi_2019}, with a peak around VMR $\sim 1 \times 10^{-5.5}$ but with an unbounded lower limit. From HST transmission spectroscopy, \citet{Barstow2017} and \citet{Pinhas2019} both retrieve a low water abundance of $1\times10^{-5}$ and $1\times10^{-4.7}$, respectively. These results are particularly significant as they are obtained with models accounting for the presence of aerosols, and therefore include their known ability to mimic a low water abundance by reducing the contrast of the water band in the WFC3 pass-band.   Lastly, a recent attempt at combining both low-resolution and high-resolution emission spectroscopy was presented by \citet{combinedhudra}, resulting in a VMR of $1\times10^{-4.1}$. The observational constraints presented above and the weak detection of water in these data inspired us to explore an additional set of models without water vapor, along with our constant VMR models with water under-abundant compared to equilibrium calculations. \\
In order to self-consistently account for Doppler shifts resulting from winds and rotation in the high-resolution spectra given that the resolution is comparable to the speeds of atmospheric motion ($\sim$km/s), 
we calculate the line-of-sight velocity for a latitude-longitude ($\theta$, $\phi$) pair at an atmospheric height of $z$ as:
\begin{multline}
       v_{LOS}(\theta,\phi) = 
       -u \sin (\theta) - v\cos(\theta) \sin(\phi) \\
       +w \cos(\theta) \cos(\phi)
    -(R_{p}+z)\Omega\sin (\theta) \cos (\phi)
\label{vloseq}
\end{multline}
where $u,v,w$ are the wind speeds in the east-west, north-south, and radial directions, respectively, and $\Omega$ is the planet's bulk rotation rate. 
We calculate simulated spectra both with and without these Doppler shifts, so that we can quantitatively evaluate how much they contribute to the observed data.
We calculate the simulated emission spectra at a higher spectral resolution ($R\sim 250,000$) than that of CRIRES data  across the same wavelength range (2.2855 -- 2.3475 $\mu$m). During the data analysis, the simulated spectra are convolved with a Gaussian kernel to match the resolving power of CRIRES ($R=100,000$).  

As the planet rotates throughout the time of observation, we calculate spectra for each exposure time. The atmospheric structure from the GCM is output every 4 degrees in phase. In order to match the more frequently sampled observed phases, we created interpolated spectra as follows. For each exposure (corresponding to some orbital phase) we take the GCM outputs from the two nearest-neighbor phases, rotate each atmosphere to the correct orientation, calculate simulated spectra from each, and then combine those two spectra, weighting linearly by how close each GCM output is to the phase of observation.  Even before this weighted average, the spectra produced from two adjacent GCM outputs differed only marginally. For the fastest rotating model, the average difference was less than $5 \% $. For the slowest rotating model, this difference was only $0.6 \%$ on average. Thus, in our interpolation process, the resultant changes to the spectra were on order of a few percent at most. 

\deleted{In addition to producing post-processed spectra from the 3-D GCM outputs, it is also instructive to compare our results against spectra produced from 1-D models of HD 209458b.  We perform comparisons against a suite of  1-D models (described in Section \ref{1Dmodels}) and choose four representative T-P profiles to show in Figure~\ref{fig: tpprof}. These four chosen representatives consist of the best fit 1-D model to the observations, two profiles that bound the temperatures produced in our GCM, and a model that approximately reproduces the average equatorial T-P profile produced by our GCM.}

\subsection{Simulated Emission Spectra} \label{sec:rt_spectra}

\begin{figure*} 
    \centering
    \includegraphics[width=\textwidth]{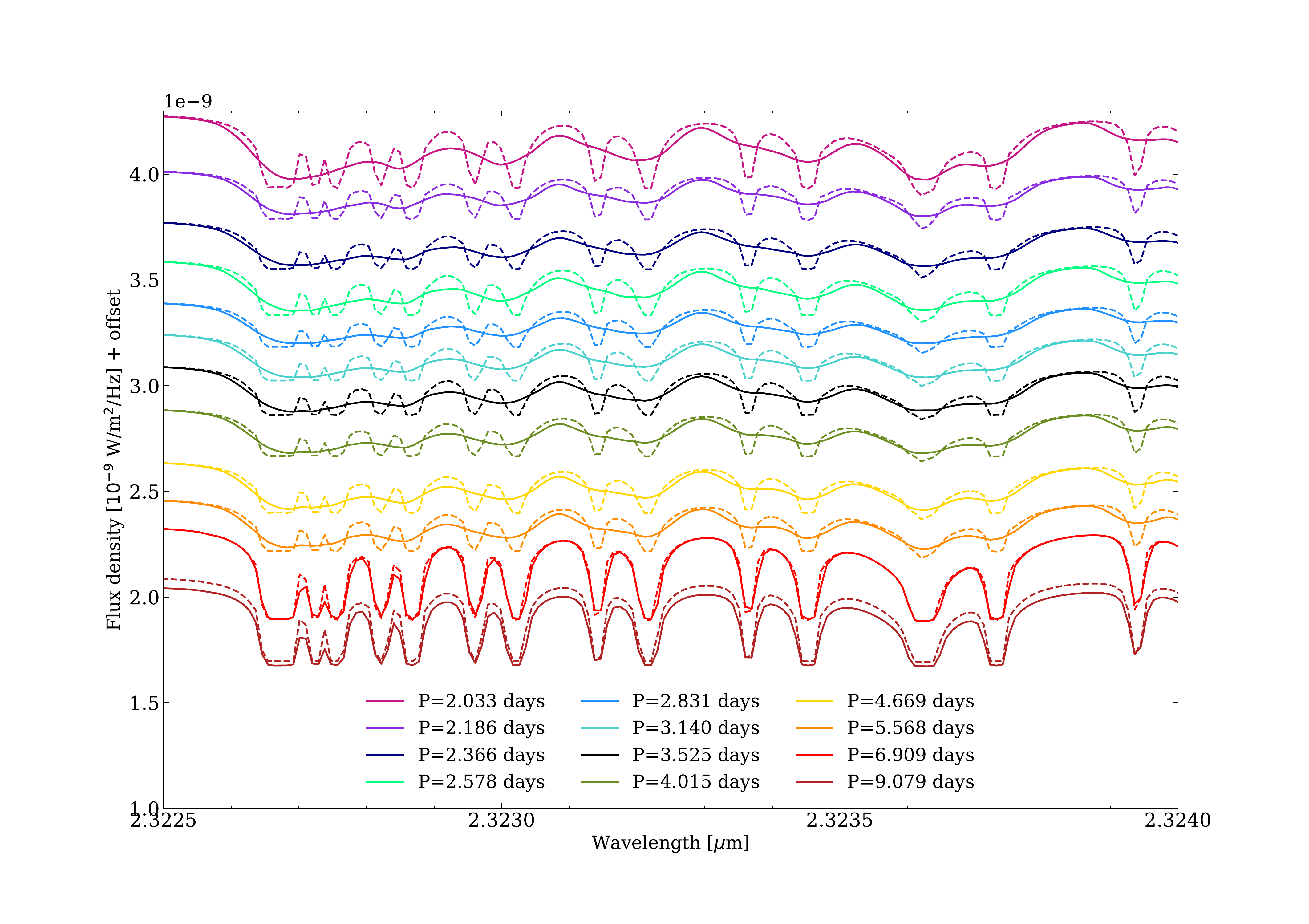}
    \caption{Simulated spectra from post-processing the atmospheric structures predicted by our GCM, color coded by the rotation rate assumed for each model (with the synchronous model in black). In these spectra we only include opacity from CO (not water; see Figure \ref{fig:synccasediffchem} for comparison) and assume local chemical equilibrium abundances. The dashed lines show spectra produced without the influence of Doppler effects while the solid lines account for shifts and broadening due to winds and rotation. The main result of the atmospheric motion is to produce significant line broadening; for most of the models the amount of broadening is similar, due to a trade-off between the contributions from winds and rotation. The two most slowly rotating models have very little broadening, due to the weak contribution from rotation, but also because of westward winds in these models.} 
    \label{fig: spec12rr}
\end{figure*}

Simulated spectra from the full set of 12 rotation models over a partial range of the total wavelength coverage for a time near the beginning of the observation (phase of $\sim$ 0.52) are shown in Figure \ref{fig: spec12rr}. For each model, versions of the spectrum with and without Doppler effects are plotted in solid and dashed lines, respectively. As expected from the discussion of their circulation patterns above, the models with the slowest rotation rates have very little Doppler broadening. In contrast, all of the other models show significant broadening, without a strong dependence on the planet's rotation rate because the faster winds in slower rotating models work to contribute to the broadening. The main notable difference between these spectra is in the relative depths of the spectral lines, which is a function of the vertical structure of these atmospheres, both thermal and as probed by the line opacities.

\begin{figure}
    \centering
    \includegraphics[width=3.6in]{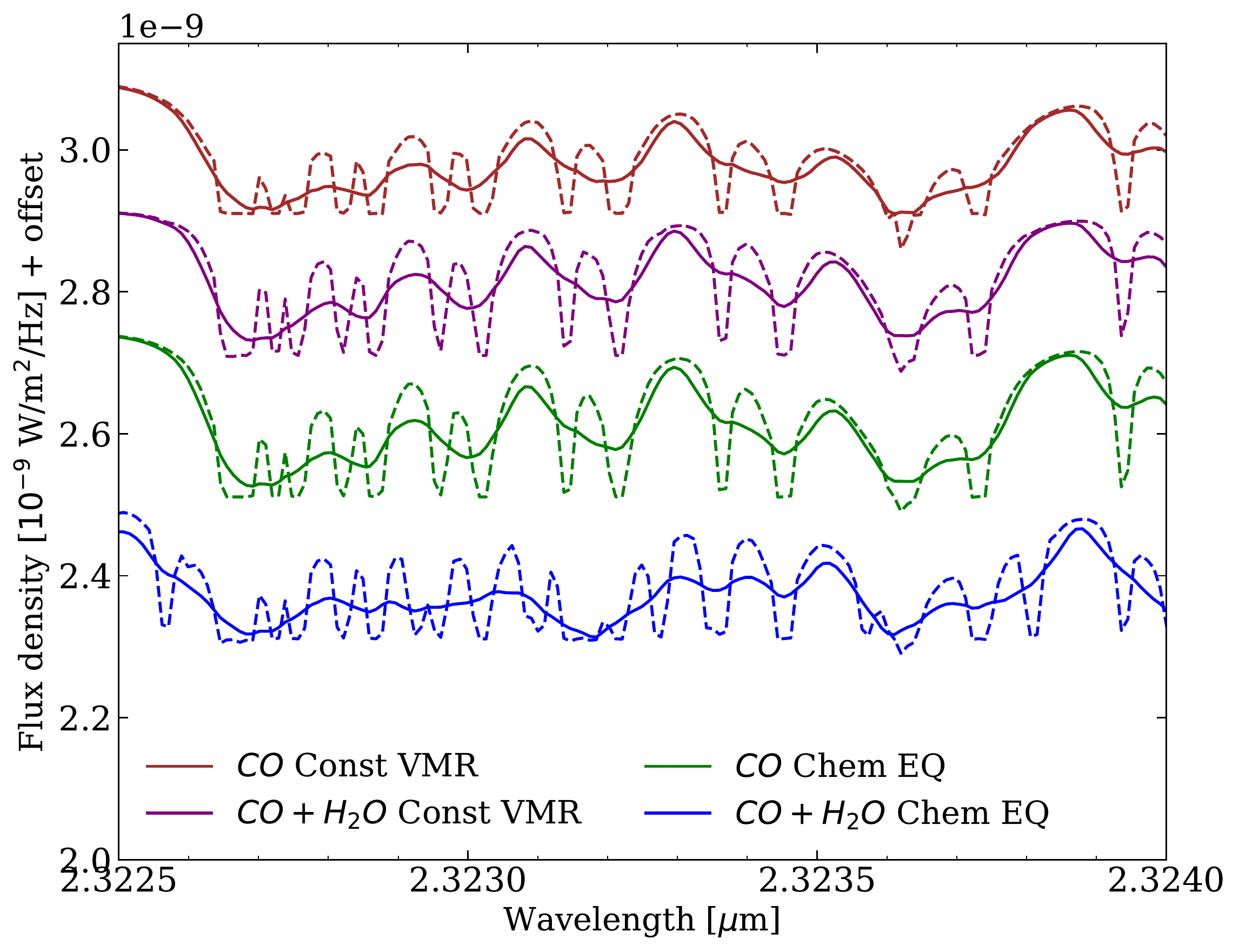}
    \caption{Simulated emission spectra, post-processed from our 3D atmospheric model, comparing the different assumptions used for the abundances of water and CO, the main sources of opacity at these wavelengths. These spectra are from the synchronously rotating model, over a fraction of the wavelength coverage of the observations; the solid and dashed spectra are produced with and without the Doppler effects of winds and rotation, respectively. The spectra produced assuming abundances determined by local chemical equilibrium and fixed to a constant value look very similar for the CO features. The assumption of local chemical equilibrium results in much more abundant water with much stronger spectral features in comparison to the constant value that best-matches previous observations.}
    \label{fig:synccasediffchem}
\end{figure}

We can investigate the relative contributions of the thermal profile and changing opacities to the depth of the spectral lines by comparing the different chemical assumptions we use in the post-processing. Figure \ref{fig:synccasediffchem} shows the differences in spectra calculated under our assumptions of chemical equilibrium abundances or constant volume mixing ratios, both with and without water included as an opacity source, for our synchronous model. The spectral features from CO appear fairly consistent for all of our assumed chemistry conditions. Over the range of pressures and temperatures that contribute to our planet's dayside emitted spectra ($P \sim 0.1-100$ mbar, $T \sim 900-1700$ K, see Figures \ref{fig:cylinmap} and \ref{fig:syncortho}), local chemical equilibrium abundances for CO at solar composition are fairly constant, at a VMR of $\sim 4\times 10^{-4}$, only slightly higher than the value we use for our constant VMR assumption. 

In contrast, the assumption of local chemical equilibrium produces significantly different water abundances than the constant VMR value we use \citep[the best-fit value from a previous 1-D analysis of these data;][]{combininglowandhigh}. For the temperature and pressure conditions probed by these emission spectra, local chemical equilibrium abundances for water have VMR $\sim10^{-3}-10^{-4}$, with the hottest and lowest pressure regions dipping down to VMR $\sim 10^{-7}$. These abundances are mostly significantly higher than our constant VMR value ($10^{-5}$), leading to much more visually apparent spectral features in Figure \ref{fig:synccasediffchem}. 
These differences will strongly influence the significance of detection in our data analysis, as discussed in Section \ref{sec:CC}.

\begin{figure}
    \centering
    \includegraphics[width=3.5in]{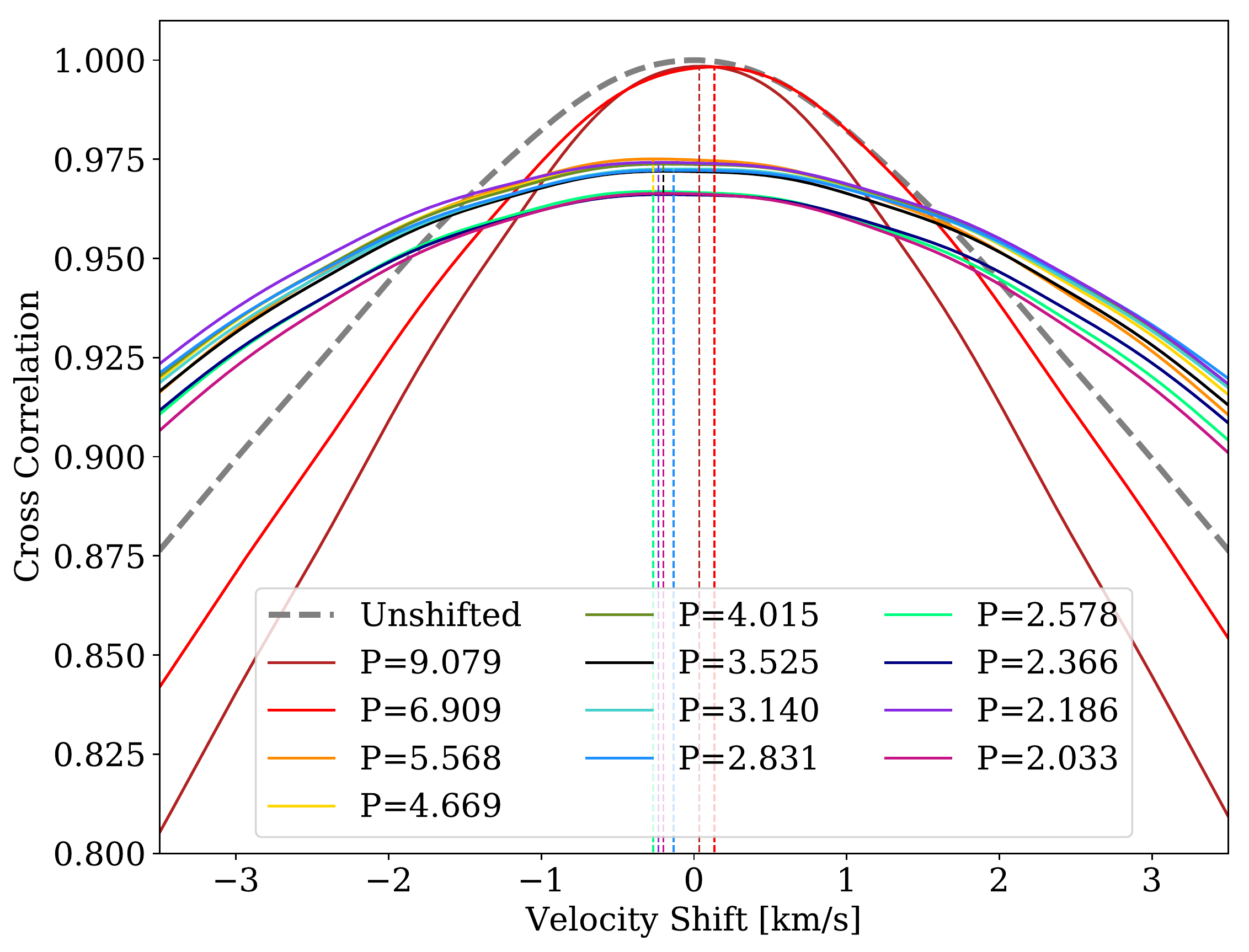}
    \caption{For each of our 12 models with different rotation rates, we cross-correlate two simulated spectra from the same model: one with Doppler effects included and one without. (The solid black line is the synchronous model.) The resulting cross correlation functions, plotted here, allow us to assess the contribution of the planet's winds and rotation to the overall Doppler shifting and broadening of the lines in the emission spectra. The gray dashed line shows the synchronous model's non-Doppler shifted spectrum, cross correlated with itself, to show the intrinsic broadening in the spectra. The dotted vertical lines mark the velocity at the peak of the cross correlation function for each model. All but the two most slowly rotating models show significant---and similar---broadening, while none of the models exhibit large net red- or blue-shifts. CRIRES allows us to fully resolve the shapes of these line profiles since its instrumental profile (approx $\sim 3$ km/s) is smaller than the FWHM of these lines. }
    \label{fig:allcc}
\end{figure}

One measure of the effect of Doppler shifting across the entire spectrum can be assessed 
by cross correlating each simulated emission spectrum with the non-Doppler shifted spectrum calculated from the same model, as shown in Figure \ref{fig:allcc}, where we have plotted these cross correlation functions for each of our 12 rotation models. 
The dashed black line shows the spectrum from the synchronous model without Doppler effects cross correlated with itself, to characterize the intrinsic width of the cross-correlation function.
The two slowest rotators have the least amount of broadening and the second slowest rotator actually has a cross-correlation function similar to the unshifted reference. All of the other rotation rates produce roughly similar levels of broadening, with only minimal net red- or blue-shifts (and no trend in the shift with rotation rate), in agreement with our previous findings in \citet{jisheng}. 

\begin{figure}
    \centering
    \includegraphics[width=3.5in]{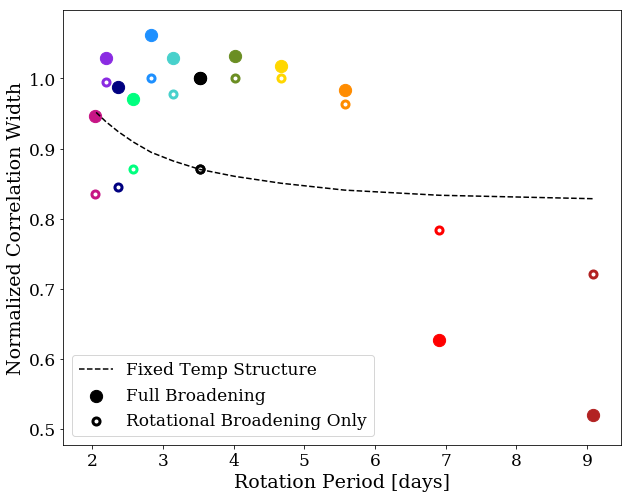}
    \caption{ Width of the cross correlation function, calculated at at a height of $80 \%$ of its maximum (shown in Figure \ref{fig:allcc}) as a function of the rotation period of the simulated planet, normalized to the synchronous model.  The filled circles correspond to spectra that have been broadened from both wind and rotation and the open circles represent spectra that have been broadened only by rotation. To produce the black dotted line, we took the temperature structure of the synchronous model and calculated the resulting broadening for each rotation rate.  For the two slowest rotating models, we find that the westward rotating winds cause the fully broadened spectra to have a smaller width than the spectra only broadened by rotation. For all of the other models, we see the addition of winds cause the resulting correlation width to increase. Because the total broadening of the line is sensitive to temperature and wind structures in addition to rotation rate, we are unable to retrieve a rotation rate from the broadening width of the spectra alone.   }
    \label{fig:FWHM}
\end{figure}

The similarity in Doppler broadening between all but the two most slowly rotating models is to be expected, from the discussions of circulation patterns above and from visual inspection of their spectra in Figure \ref{fig: spec12rr}. As a more quantitative comparison, in Figure \ref{fig:FWHM} we show the width of the cross correlation function, calculated at $80 \%$ of its maximum (shown in Figure \ref{fig:allcc}) as a function of the rotation period of the simulated planet, normalized to the synchronous model.  This width serves as a proxy to understand the degree of broadening caused by the differing sources of Doppler effects. The filled and unfilled circles correspond to spectra that have been broadened by both winds and rotation and only rotation, respectively. The scatter in the unfilled circles is a result of differences in temperature structure in the corresponding GCM. Aside from the two slowest rotating models---which exhibit westward flow, opposite of the direction of rotation---allowing the spectra to also be broadened by the winds cause the width to increase. We show the result of a single temperature structure artificially broadened at the various rotation rates with the black dashed line. The unfilled circles lie both above and below this line, meaning that the amount of broadening in the lines themselves does not allow us to constrain the rotation rate strongly. Because the total broadening of the line is sensitive to temperature and wind structures in addition to rotation rate, we are unable to retrieve a rotation rate from the broadening width of the spectra alone. 

\subsection{1D Atmospheric Models}
\added{In addition to producing post-processed spectra from the 3-D GCM outputs, it is also instructive to compare our results against spectra produced from 1-D models of HD 209458b.  We perform comparisons against a suite of previously published 1-D models (described in Section \ref{1Dmodels}) and choose four representative T-P profiles to show in Figure~\ref{fig: tpprof}. These four chosen representatives consist of the best fit 1-D model to the observations, two profiles that bound the temperatures produced in our GCM, and a model that approximately reproduces the average equatorial T-P profile produced by our GCM. }

\section{Observational Data of HD~209458b} \label{sec:data}

The data we re-analyze in this paper were originally published in \citet{schwarz}, where the full details of the observations can be found. In brief, the star HD~209458 (K=6.31 mag) was observed for a total of 17.5 hours with the CRIRES instrument on the VLT as part of the ESO program 186.C-0289 in August and September 2011. The system was observed on three separate nights, always shortly after secondary eclipse. Here we utilize only the first two nights of data, which were observed in nodding mode. We discard the third night, because this was observed in staring mode for testing purposes and shows a higher noise budget.
As explained in \citet{schwarz}, the spectra were optimally extracted via the standard ESO pipeline and then re-calibrated in wavelength using the known position of telluric lines as a reference.
Due to previously reported issues with the fourth detector of CRIRES, we chose to include only the first three detectors in our analysis.
Extracting the planetary signal from  the calibrated spectra poses a unique challenge due to the highly unequal flux ratio of the Hot Jupiter and the star. Furthermore, for ground based observations, spectral absorption lines formed in the Earth's atmosphere (telluric features) must be accounted for and are often so strong that parts of the data must be masked completely as they exhibit near-zero flux.
In order to decouple the planet's spectrum from the stellar and telluric lines, we utilize standard analysis algorithms \added{(see \citep[Section 3.2]{Brogi_2019} for a detailed description} for HRS. These are based on the principle that over the relatively short period of observations, the planetary lines are Doppler shifted by a varying amount due to the changing orbital motion of the exoplanet, while telluric and stellar lines are essentially stationary \footnote{Stellar lines do shift by $\sim100$ m s$^{-1}$ per hour of observations due to the barycentric velocity of the observer and the stellar motion around the center of mass of the system, but these are negligible compared to the change in planet's radial velocity.}. Thus, by removing the parts of our signal that do not  shift with time, we are left with the planetary spectrum. 
We apply the latest iteration of the HRS analysis described in \cite{Brogi_2019}, to which we point the reader for a step-by-step description. In short, the algorithm determines a model for the time-dependent stellar and telluric spectrum empirically from the observations, and normalizes the data by dividing out such model. The resulting data product only contains the planet spectrum, deeply embedded in the stellar photon noise at this stage.

Similarly to previous studies of atmospheric circulation from transmission spectra \citep{Brogianalysis, Flowers} we run two parallel versions of the analysis: one with the data as is (hereafter the {\sl real} data), and one containing each model spectrum injected at a small level (hereafter the {\sl injected} data), chosen to be 0.1$\times$ the nominal value. Here the nominal value is the planet's emission spectrum in units of stellar flux, i.e. scaled by a blackbody at the stellar effective temperature and multiplied by the planet-to-star surface ratio (see system parameters in Table~\ref{tab:gcm_params}). 
The exact value of the scaling factor is not important for the outcome of the analysis, as long as it is significantly smaller than the nominal value. A small scaling factor is needed to realistically simulate the effects of the analysis on each model spectrum without sensibly changing the signal content of the data. 
In order to detect the planet's emission spectrum, buried in the stellar noise at this stage, we use the standard technique in high-resolution spectra, where we cross-correlate a template spectrum---or set of templates---for the planet with the data. If the template is a good representation of the planet's spectrum, there will be a maximum cross-correlation value at velocities corresponding to the planet's orbital radial velocity during the time of observation. 

The significance of each tested model is determined as in previous work: we compute the difference between the CCF of the injected data and the CCF of the real data. This will remove the cross correlation noise and the correlation with the real planet signal, and provide us with the {\sl model} cross correlation. Note that this is different from the CCF obtained by autocorrelating the spectra, because it contains any alterations that our data analysis necessarily introduces on the planet signal while removing telluric and stellar spectra. We then compare the model CCF and the real CCF via chi-square, and we assign a significance by discriminating against a non-detection, which in our case is a flat cross correlation function (i.e. a straight line). Finally, $n$-$\sigma$ confidence intervals are determined by the region in the parameter space where the detection significance drops by $n\,\sigma$. For the full explanation of how the chi-square statistic is utilized, we refer the reader to \citet{Brogianalysis} and \citet{Flowers}.

\section{Data Analysis Results} \label{sec:CC}

We apply the cross correlation and significance test explained in Section~\ref{sec:data} to the spectra produced from our three-dimensional model, as well as to a suite of one-dimensional models for comparison. These one-dimensional models are taken from previous work and further information about them is provided in Section \ref{1Dmodels}. 

\begin{figure}
    \centering
    \includegraphics[width=3.6in]{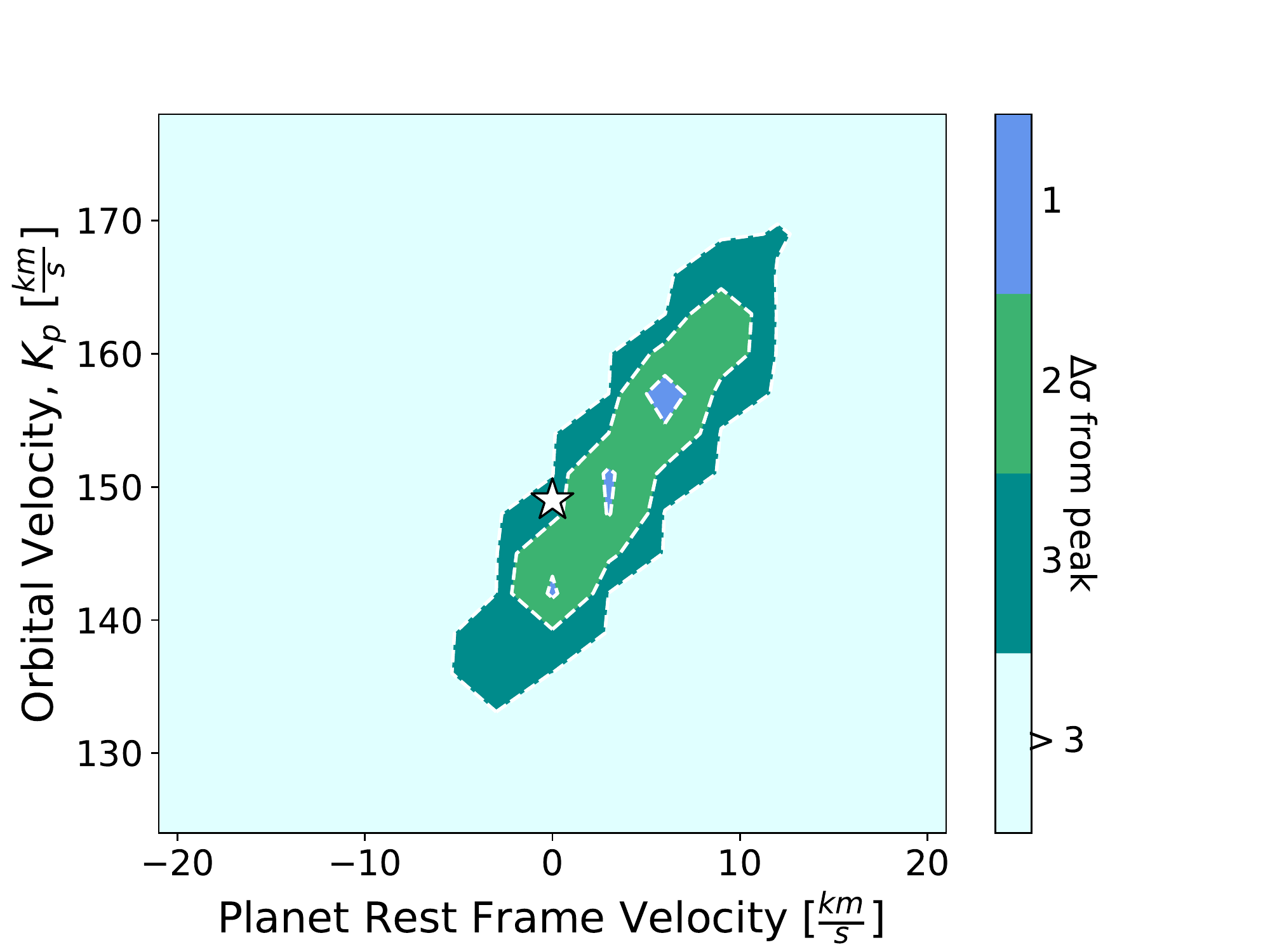}
    \caption{The significance of our detection of the planetary signal, showing 1-, 2-, and 3-$\sigma$ confidence intervals   from the peak detection (at 6.78 $\sigma$, for our spectra calculated using water and CO with constant abundances), over the velocity parameter space explored by the cross correlation fitting. The literature orbital velocity of the planet is shown as a white star, as is its expected rest frame velocity. Our analysis confidently detects the planet, at its expected velocity.} 
    \label{fig:vmrsingleh20}
\end{figure}

We find significant detection of the planet's signal over the range of template spectra tested, but our strongest detection came from the spectra produced by post-processing our three-dimensional model, as reported in Table \ref{tab:sigvalues}. In particular, we found the highest significance of detection (at 6.8 sigma) for the model that was post-processed assuming uniform volume mixing ratios for CO and water, and that included the Doppler effects from winds and rotation. Figure \ref{fig:vmrsingleh20} shows the significance of cross-correlation detection for this model, over the range of orbital and rest frame velocities included in the analysis. 
Note that these observations have a relatively small phase range and they are taken close to superior conjunction, where the planet's radial velocity curve can be approximated with a linear function of time at small signal to noise. This means that higher orbital velocities can be somewhat compensated for by allowing the planet to have a positive rest frame velocity (i.e., anomalous motion away from the observer), resulting in some inherent degeneracy between those parameters. Our detection agrees with a zero rest frame velocity for the planet and the orbital velocity reported in \citet{hd209params}.

\begin{deluxetable*}{ccccc}
\caption{Highest significance detections for the model spectra tested in this work. The highest increase in detection significance came from using a 3D atmospheric model, compared to the 697 1D models tested. Note that the best fitting 1D model exhibits a non-physical, super-adiabatic lapse rate.  \added{For detections broken down by rotation rate, see Table \ref{tab:results} in the appendix.} } \label{tab:sigvalues}
\tablehead{ \colhead{Dimensions} & \colhead{Abundances} & \colhead{Molecules Included} & \colhead{Doppler Effects On} & \colhead{Doppler Effects Off}}
\startdata
3D & Chemical equilibrium & CO & 6.49 & 6.40 \\
    & Chemical equilibrium & CO and H$_2$O & 4.22 & 3.39  \\
    & Constant volume mixing ratio & CO & 6.02 & 5.72\\
    & Constant volume mixing ratio & CO and H$_2$O & 6.87 & 6.37 \\
\hline
1D  & Constant volume mixing ratio & CO and H$_2$O & - & 5.06 \\ 
\enddata
\end{deluxetable*}

One of the main results from our analysis is this: \textit{that template spectra from our 3-D model---calculated without any fine-tuning---outperform a large suite of template spectra from one-dimensional models} (a 6.8 sigma detection significance compared to 5.1; Table \ref{tab:sigvalues}). This is evidence that the three-dimensional structure of this hot Jupiter's atmosphere leaves detectable signatures in the disk-integrated high-resolution emission spectrum of the planet. In the following sections we explore the various physical properties that could contribute to this enhanced detection and evaluate their influence.

\subsection{Comparison to 1D Models} \label{1Dmodels}

To compare our results with the modeling presented in past work, we estimated the significance of the cross correlation with two grids of models obtained with one-dimensional, plane-parallel radiative-transfer calculations. The first grid of models is described in \cite{schwarz} and consists of 704 models describing a parametric $T-p$ profile with a region at constant lapse rate ($dT/d\log(p)$) sandwiched between two isothermal regions. Pressure and temperature at the upper and lower boundaries can be changed, thus exploring a wide range of lapse rates up to $d\log(T)/d\log(p)=0.31$, which includes non-physical super-adiabatic lapse rates. Relative abundances of CO and H$_2$O are also varied in the range log(CO/H$_2$O) = 0-1.5. After excluding models with a thermal inversion layer \added{(ruled out in \citet{schwarz}} we were left with 546 models to test. Since these models were not designed to explore high abundance ratios between CO and H$_2$O, we also tested a subset of the models described in \citet{combininglowandhigh} and sampled from the low-resolution posterior retrieved by \citet{Line2016}. From that initial sample of 5,000 models we remove those models with thermal inversion and/or log(CO/H$_2$O) $< 1.5$ \added{\citep[as low CO/H$_2$O models are already included in the grid from][]{schwarz}}, resulting in 151 additional models, spanning abundance ratios up to log(CO/H$_2$O) = 3.0. All these models have a sub-adiabatic lapse rate in the range $0.05 < d\log T/d\log p < 0.08$.  The only broadening that has been applied to the 1-D model spectra arises from the pressure and thermal broadening components of the Voigt profile used to generate the spectral lines.

\begin{figure}
    \centering
    \includegraphics[width=3.6in]{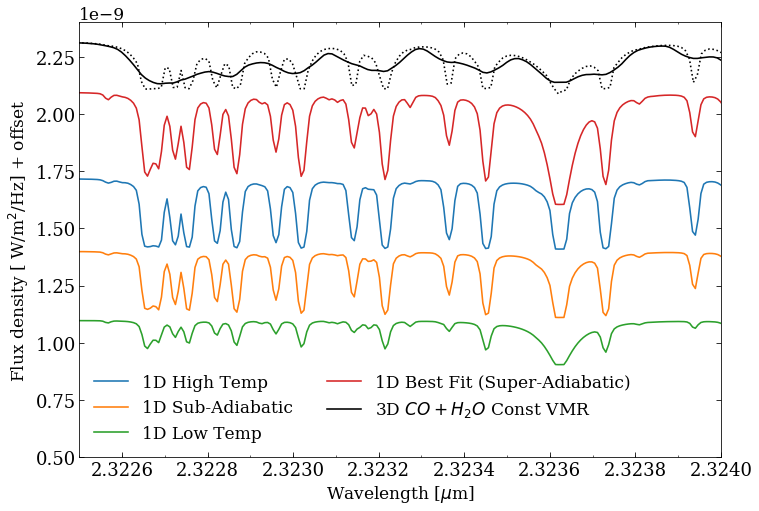}
    \caption{A comparison of the spectra produced from a 1D atmosphere with our best fitting  3D model (in black). The solid black spectrum has been broadened by Doppler effects arising from winds and rotation. These sources of broadening are not included in the dotted black spectrum or any of the 1D spectra.   All of the models appear to show the same absorption lines but the relative depths of absorption, influenced by the underlying temperature structure and chemical abundances, changes with each model. These variances in relative depth and line shape result in a range of significance of detection when cross correlated with the data.  }
    \label{fig:1dspectra}
\end{figure}

Of the 697 one-dimensional models tested, the highest measured significance is 5.06$\sigma$, with only 14 models reaching a significance value greater than 4$\sigma$. These are models with a steep lapse rate ($0.13 < d\log T/d\log p < 0.31$) and an abundance ratio of 10-30 between CO and H$_2$O. Thus, the vast majority of the 1-D models return a significance below the threshold of detection (usually set at 4$\sigma$ for these HRS observations), and consistent with the tentative detection reported in \cite{schwarz}. 
We note that the temperature-pressure profiles explored in the set of 1-D models encompasses the range realized in our 3-D model (Figure \ref{fig: tpprof}).  This implies that the deficiency in the 1-D models is not that they didn't include the \textit{appropriate } physical conditions of the planet, but rather that those conditions are inherently, and observably, three-dimensional. In Figure \ref{fig:1dspectra}, we show a subset of the spectra produced from the 1D models and spectra from our best fitting 3D model. All the spectra shown seem to show the same absorption lines, yet still result in a range of detection strengths. The subtleties in spectral line shapes and relative depths are not adequately captured by the 1D models. 

\subsection{Influence of Temperature Structure}

As Table \ref{tab:sigvalues} hints at, and as we will discuss in subsequent sections, the improvement in detection from using the 3-D models over the 1-D models is not primarily due to the chemical or velocity structure of the atmosphere, as those influences on the spectrum only give marginal improvements in the significance of detection. Instead, we find that the contribution from multiple regions of the planet, with different thermal structures, is a much better match to the observed data than a representation of the planet with a single thermal profile. Whether the influence of spatial inhomogeneity is intrinsically within \textit{all} HRS emission observations requires further study, but for this particular planet we find it to be the case. Recent complementary work by \citet{Taylor2020} predicts that James Webb Space Telescope observations may similarly contain inherent signatures of multiple thermal regions, although whether this inhomogeneity will be measurable or not depends on wavelength coverage and signal-to-noise.
\subsection{Influence of Chemical Structure}
Table \ref{tab:sigvalues} shows that for models with CO alone the assumption of abundances that follow local chemical equilibrium is slightly preferred over using the best-fit value from a previous analysis of these data \citep{Brogianalysis}. However, as discussed in Section \ref{sec:rt_spectra}, local chemical equilibrium does not predict strong variations in the abundance of CO throughout the atmosphere, meaning that the improvement of signal does not come from any significant chemical heterogeneity influencing the disk-integrated spectra, but rather from an abundance slightly closer to reality. It may be the case that by capturing the inherent \textit{thermal} inhomogeneity of the atmosphere, we can more accurately find the correct chemical abundances \citep[][]{Taylor2020}.

In contrast to our results for CO, Table \ref{tab:sigvalues} shows a strong decrease in the significance of planet detection when using chemical equilibrium values for water. The data prefer depleted abundances for water; Section \ref{sec:rt_spectra} and the discussion surrounding Figure \ref{fig:synccasediffchem} demonstrate that water at equilibrium values would result in large spectral features that are not apparent in the data, according to our analysis. It is noteworthy that the data are not suggesting a complete lack of water; the very low water abundance used in calculating the spectra with constant VMR does improve the planet detection over the comparable CO-only model.

A full gridded analysis of varying chemical abundances is outside the scope of this work. Even without considering a full grid,these results show that the 3-D chemical structure of the atmosphere contributes to our enhanced detection, compared to 1-D models, insofar as it seems to slightly more robustly predict the abundance of CO in the atmosphere. Notably, we find that the data prefer a water abundance that is orders of magnitude depleted below chemical equilibrium values.

\subsection{Influence of Atmospheric Doppler Effects}

In addition to predicting the 3-D temperature structure of the planet's atmosphere, our GCM also predicts the wind vectors throughout, all of which are influenced by the rotation rate assumed for the planet. Here we examine how the Doppler shifts and broadening due to winds and rotation in our simulated spectra may contribute to our enhanced detection of the planet's signal over the 1-D models that do not include this additional physics, and whether the data can help to empirically constrain the planet's wind speeds and rotation rate (generally assumed to be synchronous with its orbit). 

In Table \ref{tab:sigvalues} we report that including the spectral line shifting and broadening from winds and rotation does enhance our detection of the planet, but with only a minor increase in significance over the spectra without Doppler effects. As discussed and shown above in Figure \ref{fig:allcc}, the main influence of the Doppler effects (for most of the models) is to broaden the spectral lines, since both winds and rotation contribute similar symmetric velocity patterns. Thus we expect the main contribution to the increased detection is that the planet's actual spectrum does contain some significant broadening from winds and rotation.

\begin{figure*}
    \centering
    \includegraphics[width=\textwidth]{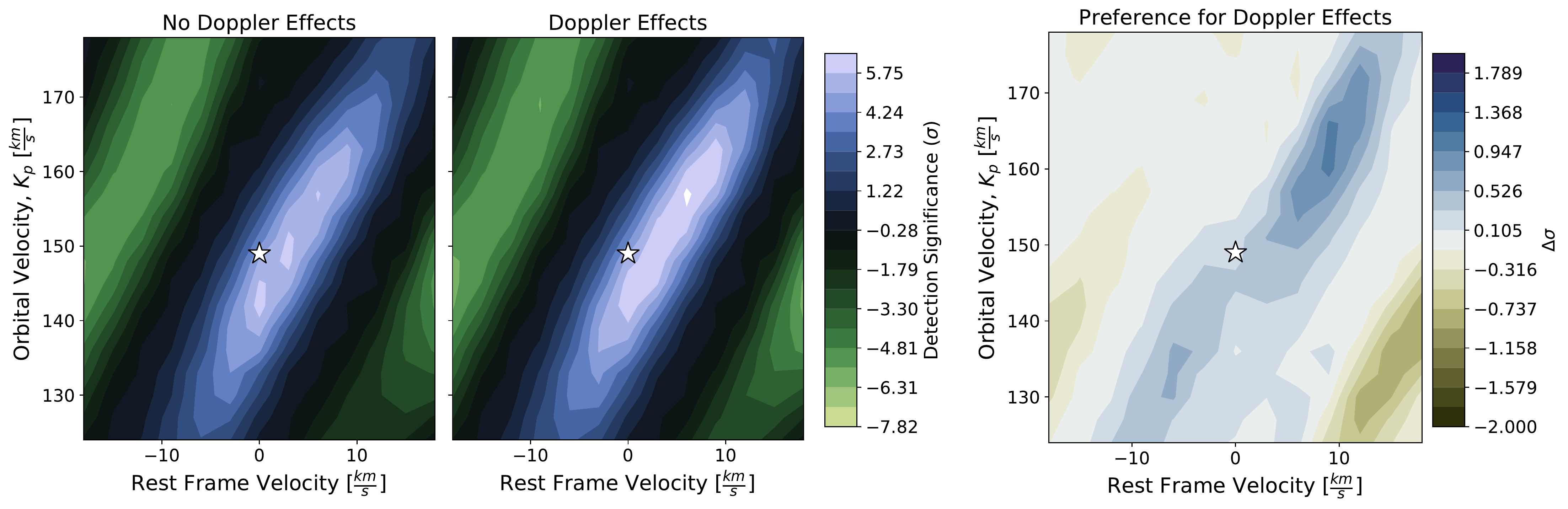}
    \caption{A comparison between our significance of planet detection with and without Doppler effects included in our  best-fit simulated spectra (\textit{left} and \textit{middle} plots), shown as a function the planet's assumed orbital velocity and its rest frame velocity (which should be zero unless there is anomalous motion). The \textit{right} plot shows the difference in significance caused by including the Doppler effects in our analysis. While there is a slight increase in detection significance, this does not correspond to any net shift in velocity space, indicating that it is largely due to the line broadening rather than any shifting. }
    \label{fig:dopoffkpev}
\end{figure*}

Even with a symmetric velocity field, an uneven brightness pattern across the planet can result in the red- or blue-shifted side of the planet contributing more emission to the disk-integrated spectrum, resulting in a net Doppler shift \citep{jisheng}. Figure \ref{fig:allcc} has small net Doppler shifts for the models. Depending on the precision of the data, this could result a small anomalous radial velocity of the planet if not included in the analysis. In order to test whether a net Doppler shift contributes in any significant way to our detection, in Figure \ref{fig:dopoffkpev} we plot the models' significance of detection in velocity space, comparing the spectra with and without the Doppler effects included. While we see an overall increase in detection significance with the Doppler effects included, there is no very noticeable shift in velocity space between the models with and without them. This agrees with our discussion above, that the main improvement in significance comes from the broadening of the lines, rather than any net Doppler shift.

\subsubsection{Constraints on rotation and winds?}
\begin{figure}
    \centering
    \includegraphics[width=3.6in]{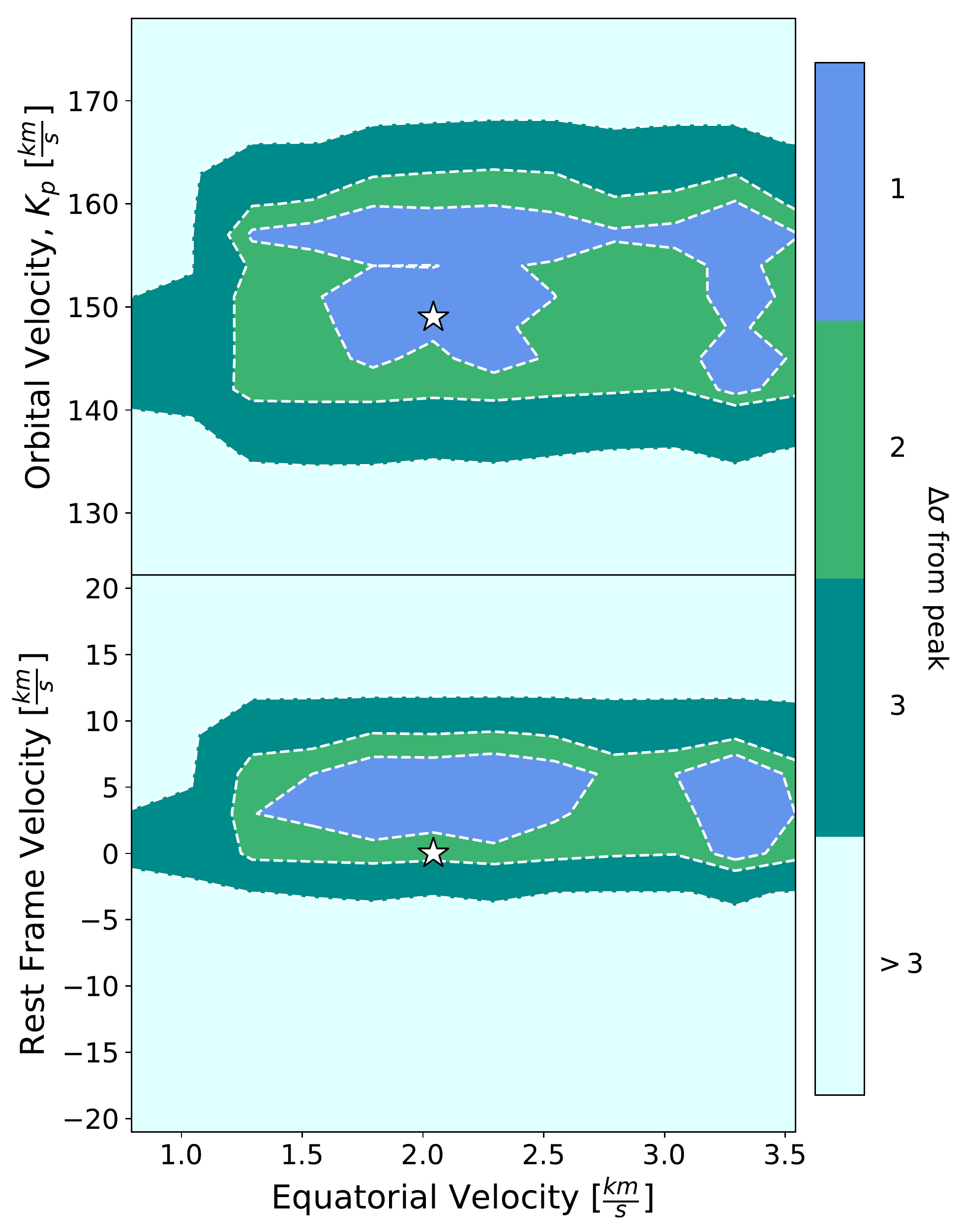}
    \caption{\replaced{Contours of detection significance for}{Confidence intervals from} cross-correlation between the data and our 3D models with constant volume mixing ratios of CO and water, and Doppler effects included. Similar to Figure \ref{fig:vmrsingleh20},  the white star marks literature values and equatorial velocity for synchronous rotation. Here, the two plots \replaced{compare the significance of detection }{show the  1-, 2-, and 3-$\sigma$ confidence intervals}for models with different rotation rates as a function of orbital velocity (\textit{top}) and rest frame velocity (\textit{bottom}). The data have a slight aversion to the two most slowly rotating models (low values of equatorial velocity), but otherwise the temperature structures and wind patterns of all other models are roughly equally well allowed by the data. }
    \label{fig:combinedvmrh20}
\end{figure}

As part of this investigation, we wanted to see what constraint, if any, could be placed on the rotation rate or wind speeds for HD~209458b. In Figure \ref{fig:combinedvmrh20} we show how the significance of detection depends on which rotation rate we use in our 3-D model of the planet (plotted here as the planet's equatorial velocity). The significance of detection is largely insensitive to the planet's rotation rate, aside from the two most slowly rotating models being slightly disfavored (and those are also inconsistent with thermal phase curve data; see Figure \ref{fig:pcurve} and discussion). Our small improvement in detection from including Doppler effects, combined with the strong similarity in Doppler broadening for all but the slowest models (Figure \ref{fig:allcc}), makes this result unsurprising. 

However, it is a valuable result to determine that the amount of Doppler broadening for models across a wide range of rotation rates is so similar (quantified in Figure \ref{fig:FWHM}). 
It indicates that we cannot constrain rotation rates as well as we might think from rotational broadening alone; the winds are faster in the more slowly rotating models and their predominantly eastward direction lets them compensate for the weaker rotational broadening.
Although our particular analysis is only for observations around one particular orbital phase, the eastward wind pattern extends around the whole globe and so we expect the same behavior regardless of orbital phase.
This is the same general behavior previously reported for high-resolution transmission spectra in \citet{Flowers}; we have now shown that emission spectra are subject to this inherent physical uncertainty as well.

\section{Conclusions and Summary} \label{sec:conclusion}

In this project, we combined state of the art observational and modeling techniques to obtain \replaced{a result stronger}{a higher significance detection} than could be achieved with either of these techniques alone. We ran a 3D atmospheric model for the hot Jupiter, HD 209458b, for a range of rotation rates. We post-processed the resulting atmospheric structures in a geometrically correct way to generate template spectra. We then cross correlated the synthetic spectra with previously published data for this planet from CRIRES/VLT and detected the planet at a greater significance than a whole suite of 1D models.  We explored why the 3D models were a strong improvement over the 1D models by looking at properties such as temperature and chemical structure and Doppler shifts from winds and rotation.  Our main findings are summarized as follows:
\begin{itemize}
    \item High resolution emission spectra are sensitive to the 3D structure of the atmosphere, at least for these data of this particular hot Jupiter. 
    \item One dimensional models, despite covering the same range in temperature and pressure, returned detections that were at best $\sim 1.8
    \sigma$ lower than our best fit from 3D models.
    \item In terms of detection significance, the primary improvement is from the use of a 3D temperature structure, with secondary improvements related to the chemistry and Doppler effects.  
    \item Doppler shifts are present in the high resolution spectra, but are unable to offer strong constraints for wind speed or rotation rate. We have shown that the widths of the spectral lines cannot be directly related to the planet's rotation rate alone.
    \item Our analysis detects water in these high resolution spectra of HD 209458b, but at a significantly depleted value \added{compared to the solar chemical equilibrium abundance} . 
\end{itemize}

High resolution spectroscopy enables detailed characterization of exoplanets. It is becoming increasingly clear that the three-dimensional nature of planets and their atmospheric dynamics influence high resolution spectra. Looking toward the upcoming era of high resolution spectrographs on Extremely Large Telescopes, we eagerly await what detailed atmospheric characterizations will be possible.  

\acknowledgments
This research was supported in part by NASA Astrophysics Theory Program grant NNX17AG25G and the Heising-Simons Foundation. MB acknowledges support from the UK Science and Technology Facilities Council (STFC) research grant ST/S000631/1. \added{We thank the referee for their constructive feedback which helped to improve the clarity of this paper.}

\bibliographystyle{aasjournal}
\bibliography{bib.bib}

\section{Appendix} \label{sec:appendix}

Here we present the GCM results for our 12 models of HD~209458b with different rotation rates, showing the temperature and wind structures of the model atmospheres.

\begin{figure*}
    \centering
    \includegraphics[width=8in]{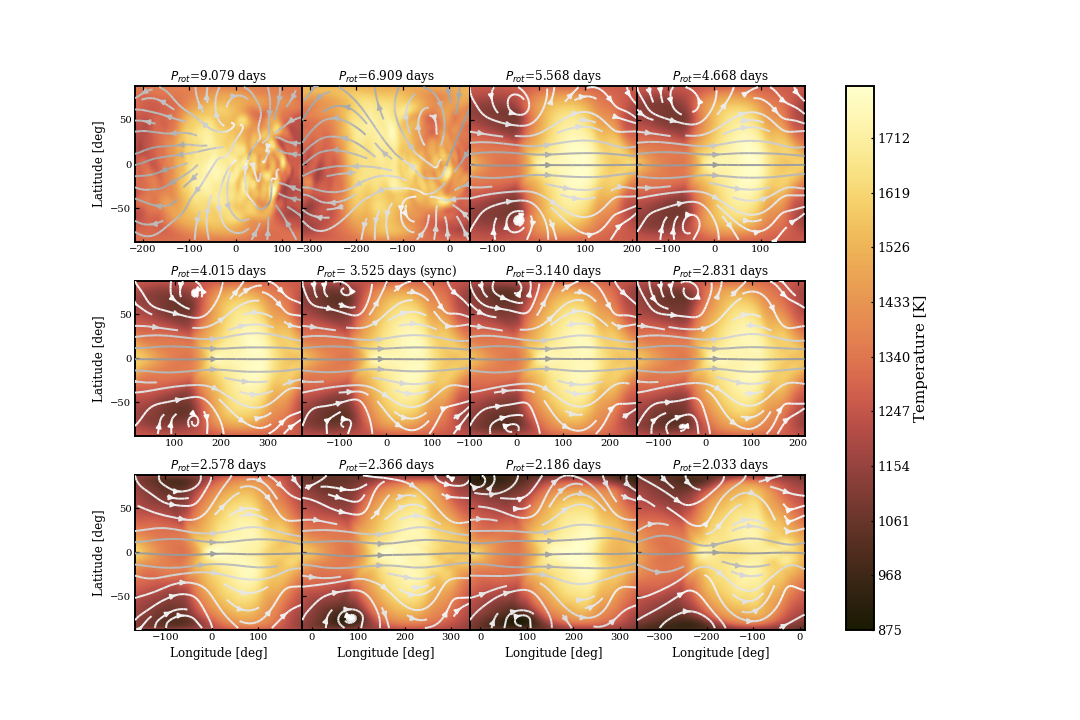}
    \caption{Temperature \added{and wind} structure at the infrared photosphere \added{(P=65 mbar)} for all 12 GCMs. In each case the orientation of the map is such that the substellar point is in the center of the plot. While most models show a temperature structure influenced by the standard hot Jupiter eastward equatorial jet, the two most slowly rotating models have disrupted circulation patterns and instead have their hottest regions shifted slightly westward of the substellar point.}
    \label{fig:alltemps}
\end{figure*}

\begin{figure*}
    \centering
    \includegraphics[width=8in]{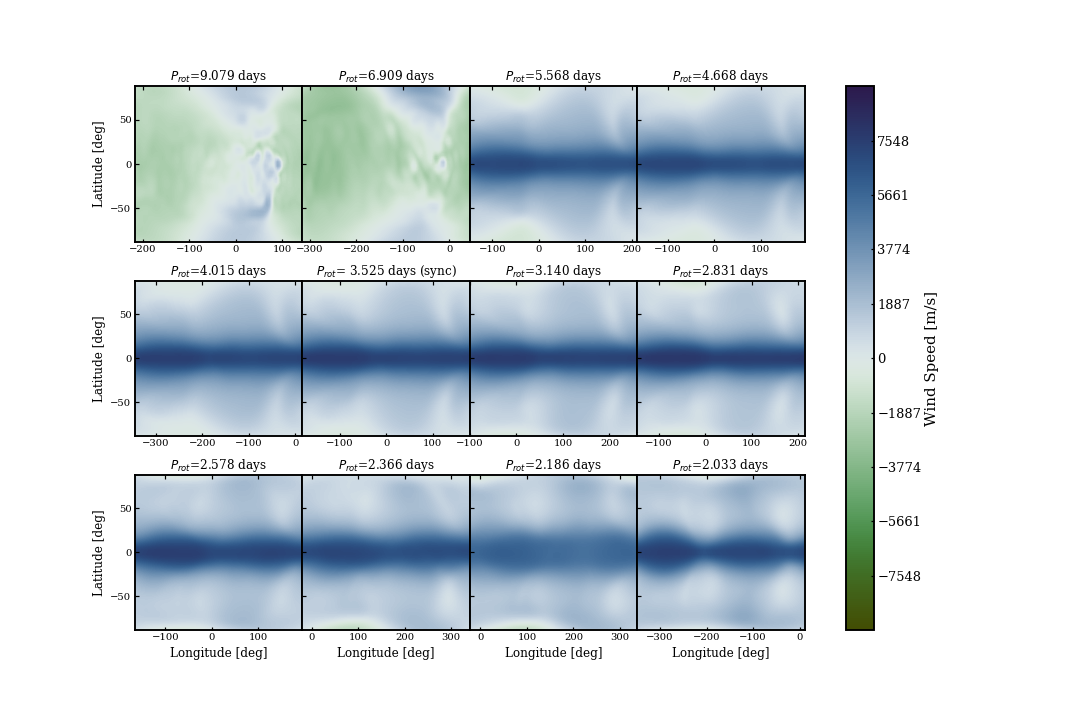}
    \caption{Maps of the winds in the east-west direction (with eastward defined as positive values) at the infrared photosphere \added{(P=65 mbar)} of the planet, for our full suite of General Circulation Models. Each map is oriented to be centered on the substellar point. Most models show the standard eastward equatorial jet, but the two most slowly rotating models have no coherent equatorial jet and instead have westward flow near the substellar point and across most of the planet.}
    \label{fig:windgrid}
\end{figure*}

\begin{figure*}
    \centering
    \includegraphics[width=7in]{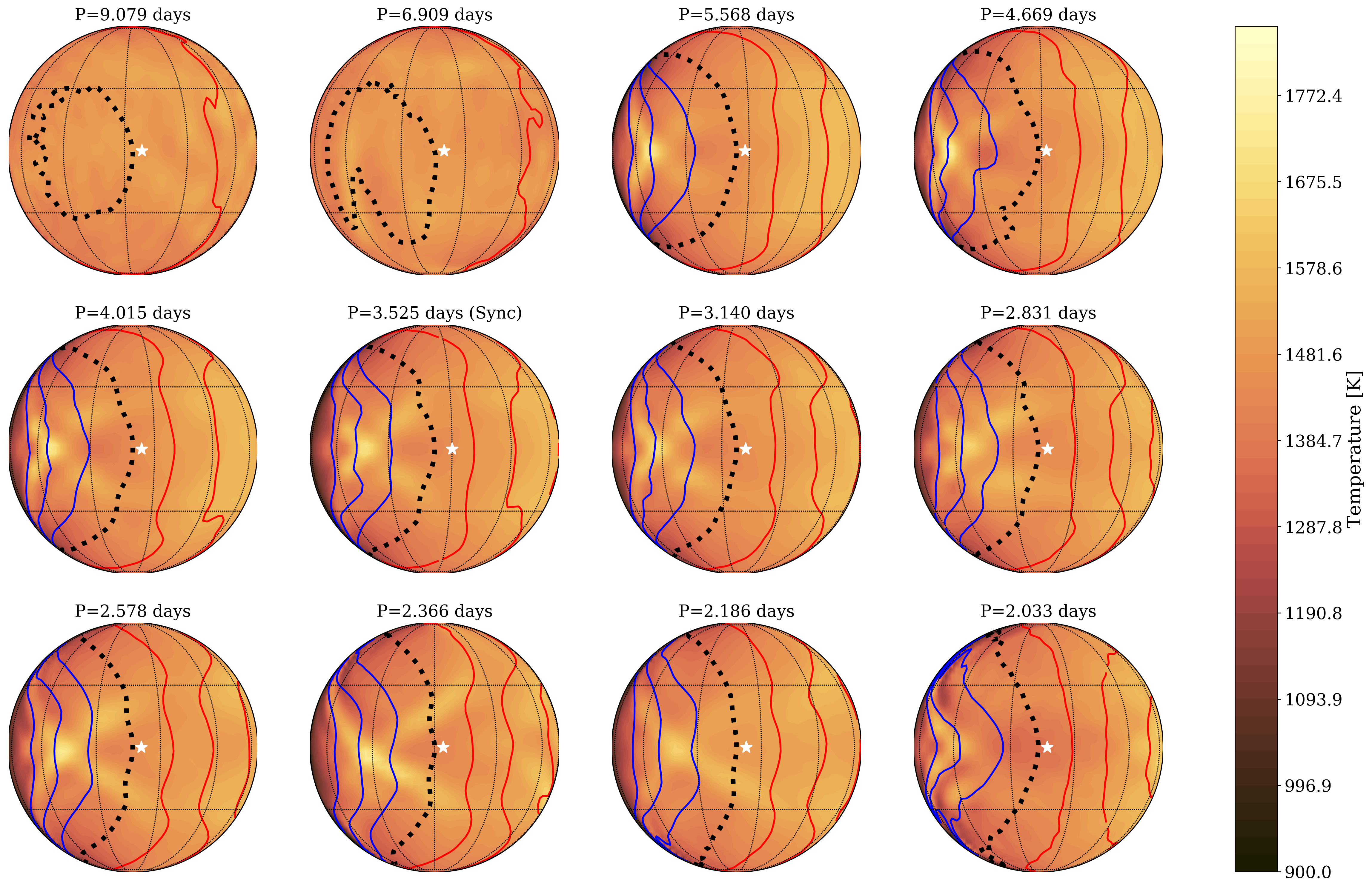}

    \caption{Orthographic projections of the temperature structure for 12 different rotation rates shown at the atmospheric level responsible for the the strongest absorption lines in the post-processed spectra, orientated such that the subobserver point is centered. Red and blue contours show constant line of sight velocities at 2, 4, and 6 km/s. The black dotted contour shows 0 km/s line of sight and the white star shows the substellar point.     Aside from the models experiencing a disrupted flow pattern---corresponding to the slowest two rotation rates---the temperature structure and circulation pattern are fairly similar over different rotation rates. Even though the rotation rate of the planet is increasing, the winds are decreasing in strength in such as way that results in similar line of sight velocity patterns across the models. We also see that these line of sight velocity patterns are not symmetric and are influenced by the underlying wind structure. }
    \label{fig:orthogrid}
\end{figure*}
\begin{figure*}
    \centering
    \includegraphics[width=7in]{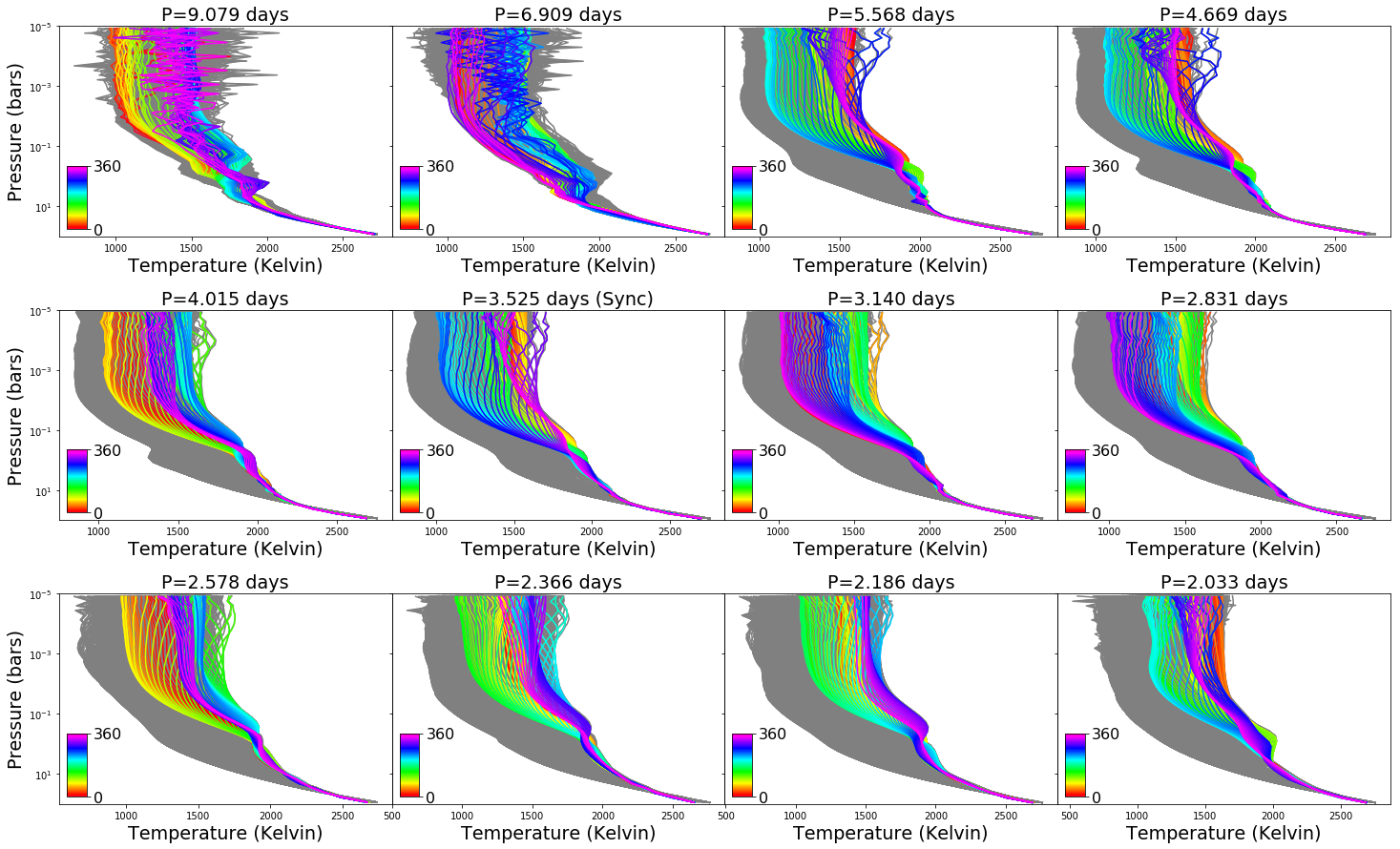}
    \caption{Temperature pressure profiles for the suite of models examined. Similar to Figure 5, the grey profiles are from the entire planet. The rainbow lines show equatorial profiles. Since these rotation rates are not equal to the period, the subobserver and substellar longitudes are not constant. We report the subobserver longitudes, starting with the slowest rotator as: [210, 140, 320, 330, 170, 350, 100, 70, 170, 240, 260, 310] degrees. The numerical noise, seen most prominently in the upper atmospheres of the two slowest rotators, has little effect on the resulting spectra.} 
    \label{fig:alltp}
\end{figure*}

\begin{deluxetable}{cccccc} 
\caption{Peak Detections for All Rotation Rates}\label{tab:results}
\tablehead{
	 & \multicolumn{2}{c}{Chem EQ} &  \multicolumn{2}{c}{Constant VMR} \\
	 \colhead{Period (days)} &\colhead{CO only} & \colhead{CO and H$_{2}$O} & \colhead{CO only} & \colhead{CO and H$_{2}$O} } 

\startdata
9.079         & 5.57/5.57            & 3.01/2.95  & 5.51/5.51        & 5.31/5.29  &  \\
6.909         & 5.74/5.70            & 3.00/2.92  & 5.54/5.50        & 5.54/5.47  &  \\
5.568         & 6.03/5.97            & 3.93/3.28  & 5.37/5.29        & 6.50/6.23  &  \\
4.669         & 6.05/6.02            & 3.91/3.26  & 5.20/5.13        & 6.59/6.28  &  \\
4.015         & 6.01/6.00            & 3.92/3.16  & 5.33/5.12        & 6.73/6.23  &  \\
3.525 (sync)  & 6.19/6.04            & 4.02/3.24  & 5.40/5.20        & 6.74/6.15  &  \\
3.140         & 6.16/6.04            & 4.04/3.27  & 5.39/5.09        & 6.77/6.21  &  \\
2.831         & 6.01/5.95            & 4.05/3.33  & 5.25/4.96        & 6.72/6.15  &  \\
2.578         & 6.12/6.00            & \textbf{4.22}/3.37  & 5.46/5.25        & 6.55/6.02  &  \\
2.336         & 6.10/5.89            & 4.21/3.31  & 5.51/5.17        & 6.62/5.88  &  \\
2.186         & \textbf{6.49}/\textbf{6.40}            & 4.19/\textbf{3.39}  & 5.47/5.43        & \textbf{6.87}/\textbf{6.37}  &  \\
2.033         & 6.43/6.05            & 4.18/3.18  & \textbf{6.03}/\textbf{5.72}        & 6.54/5.81 &  \\
\enddata
\tablecomments{\added{Peak detection for every 3D model examined with Doppler effects considered (first entry) and without (second entry). The highest detection for each chemistry and Doppler setting across all rotation rates is bolded and reported in Table \ref{tab:sigvalues}. While the highest significance detections come from the more quickly rotating models, these values are only minimally above those for the synchronous model.} }
\end{deluxetable}

\listofchanges
%

\end{document}